\documentclass[aps,prb,twocolumn,amsmath,amssymb,superscriptaddress,floatfix]{revtex4-1}

\usepackage{amssymb}
\usepackage{color}
\usepackage{graphicx}
\usepackage{mathrsfs}
\usepackage[unicode=true,pdfusetitle,
 bookmarks=false,colorlinks=true,citecolor=blue,urlcolor=blue,linkcolor=red]{hyperref}

\begin{document}

\title{
Global anomalies on the surface of fermionic  symmetry-protected topological phases in (3+1) dimensions 
   }

\author{Chang-Tse Hsieh} 
\affiliation{
Department of Physics, University of Illinois
at Urbana-Champaign, 
1110 West Green St, Urbana IL 61801
            }

\author{Gil Young Cho}
\affiliation{
Department of Physics, University of Illinois
at Urbana-Champaign, 
1110 West Green St, Urbana IL 61801
            }

\author{Shinsei Ryu}
\affiliation{
Department of Physics, University of Illinois
at Urbana-Champaign, 
1110 West Green St, Urbana IL 61801
            }

\date{\today}

\begin{abstract}
Quantum anomalies, breakdown of classical symmetries by quantum effects,  
provide a sharp definition of symmetry protected topological phases.
In particular, they can diagnose interaction effects on the non-interacting classification of fermionic symmetry protected topological phases.
In this paper, we identify quantum anomalies
in two kinds of (3+1)d fermionic symmetry protected topological phases:
(i) topological insulators protected by CP (charge conjugation $\times$ reflection) and electromagnetic $\mathrm{U}(1)$ symmetries,
and 
(ii) topological superconductors protected by reflection symmetry.
For the first example, which is related to, by CPT-theorem, time-reversal symmetric topological insulators,  
we show that the CP-projected partition function of the surface theory is not invariant under large $\mathrm{U}(1)$ gauge transformations,
but picks up an anomalous sign, signaling a $\mathbb{Z}_2$ topological classification.  
Similarly, 
for the second example, 
which is related to, by CPT-theorem, time-reversal symmetric topological superconductors,  
we discuss the invariance/non-invariance of the partition function of the surface theory, 
defined on the three-torus and its descendants generated by the orientifold projection, 
under large diffeomorphisms (coordinate transformations).
The connection to the collapse of the non-interacting classification by an integer ($\mathbb{Z}$) 
to $\mathbb{Z}_{16}$, in the presence of interactions,  is discussed.   
\end{abstract}


\maketitle

\tableofcontents

\section{Introduction}
\label{Introduction}

Gapped states of quantum matter can be topologically distinguished 
by asking if given ground states can be adiabatically connected to each other.
Topologically equivalent phases can continuously be connected in the phase diagram 
(``theory space'' in the language of the renormalization group)
with no gapless phase boundary nor quantum critical point separating them.  
On the other hand, topologically distinct phases are always separated by intervening gapless 
phases or quantum critical points. 

Quite often, it is meaningful to discuss theory space in the presence of a set of symmetries.
Symmetries may prohibit the appearance of some phases distinct topologically from trivial phases, 
while they may also create a new topological distinction among quantum phases. 
Symmetry-protected topological (SPT) phases of matter correspond to the latter case --
they are topologically equivalent to trivial states of matter (such as an ionic insulator)
in the absence of symmetries, while once a certain set of symmetries are imposed, 
they cannot be adiabatically connected to topologically trivial phases. 
For partial references for recent works on symmetry protected topological phases, see
Refs.\ 
\onlinecite{
HasanKane2010, QiZhang2011, HasanMoore2011, Schnyder2008, SRFLnewJphys, Kitaev2009, 
Chen2013, Chen2012, Gu2014, Lu2012, LevinGu12}.

As different SPT phases respect the same symmetry, 
they cannot be characterized by
the Landau-Ginzburg-Wilson paradigm based on spontaneous symmetry breaking. 
Hence one needs to look for an alternative organizing principle.
Quantum anomalies, an intricate form of symmetry breaking caused by quantum effects, 
have been proved to be useful in this context. 
Already in Laughlin's gauge argument, 
a topological charge pumping process, i.e., 
the non-invariance of the system's ground state under large $\mathrm{U}(1)$ gauge transformations,
was used to establish the stability of the quantum Hall states against interactions and disorder.\cite{Laughlin1981}
For SPT phases, see recent works in Refs.\ 
\onlinecite{Cho2014,Ringel2013, Cappelli13, RyuMooreLudwig2012, Koch-Janusz2014, Kapustin2014a,Kapustin2014b, Kapustin2014c,Wang2013, Wang2014, You2015}.

In this paper, basing on our previous works on (2+1)d SPT phases,
\cite{Ryu2012, Sule13, Hsieh2014a, Cho2015}
the surface states of (3+1)d fermionic SPT phases are studied
from the perspective of global quantum anomalies. 
We will discuss two examples:
(i)
the Dirac fermion surface state of (3+1)d bulk $\mathrm{CP}$ symmetric topological insulators (TIs)
and 
(ii)
the Majorana fermion surface state of (3+1)d bulk reflection symmetric crystalline topological superconductors (TSCs).  
The bulk phase of the first example is a fermionic SPT phase protected by 
electromagnetic $\mathrm{U}(1)$ and CP [product of charge conjugation and mirror reflection (parity)] symmetries.
This example is CPT-conjugate to a (3+1)d time-reversal symmetric TI (class AII),
and characterized by a $\mathbb{Z}_2$ topological number. \cite{Hsieh2014b} 
The bulk phase of the second example is a fermionic SPT phase protected by
fermion number parity and reflection (parity) symmetry.
It belongs to symmetry class D+R$_{+}$ crystalline TSCs
\footnote{
The subscript ''+'' in symmetry class D+R$_{+}$
indicates that the two symmetry operations, 
charge-conjugation (or particle-hole) and reflection symmetries, 
of the single-particle Hamiltonians in this symmetry class
commute with each other. 
}
in Refs.\ \onlinecite{Chiu13, Morimoto13, Shiozaki14}.
This example is CPT-conjugate to a (3+1)d time-reversal symmetric TSC (class DIII).
\cite{Hsieh2014b}
At non-interacting level,
class D+R$_{+}$ crystalline TSCs in (3+1)d are characterized by an integer ($\mathbb{Z}$) topological number
(the Mirror Chern number),
similar to their CPT partner, class DIII TSCs.
\footnote{
While we are not to be restricted to relativistic systems in condensed matter physics,  
some universal physical properties of general, non-relativistic systems in the long wavelength limit, 
such as the band topology or the electromagnetic responses, 
are often encoded in topological field theories.
Since topological, these theories respect the Lorentz symmetry, which guarantees the CPT invariance.
In addition, from the perspective of topological classification of states of matter, 
classifying SPT phases of non-interacting fermion systems, for example, 
can be done solely in terms of Dirac operators with symmetry restrictions.
Since a Dirac Hamiltonian has a CPT invariant form, 
we expect to obtain the same classification for all CPT equivalent systems, e.g., CP-protected TIs to class AII TIs, and classes D+R$_{+}$ TSCs to class DIII TSCs discussed here.
}
On the other hand, 
a number of recent works 
showed that the integral non-interacting classification of class DIII TSCs 
breaks down to $\mathbb{Z}_{16}$ once interactions are included. 
\cite{Metlitski2014, Fidkowski2013, Kapustin2014c, You2014, KitaevUnpublished, WangSenthil2014, Senthil2014}
Such collapses have been also reported in one and two spatial dimensions.
\cite{Fidkowski2010,Fidkowski2011,Turner2011,Tang2012,Qi2013, Ryu2012, Yao2012,GuLevin2014} 


Similar to our previous work, \cite{Hsieh2014a} 
we enforce $\mathrm{CP}$ or reflection symmetry on the surface theories by taking an orientifold projection.
\cite{Callan1987, PolchinskiCai1988, Horava1989, Angelantonj02, Sagnotti1988,Dai1989} 
(See also recent discussion in Refs.\ \onlinecite{Chen2014, Kapustin2014a}.)
In the first example, the resulting projected theory is then shown to have global $\mathrm{U}(1)$ gauge anomaly.
That is, the partition function of the projected theory picks up a phase under large $\mathrm{U}(1)$ gauge transformations. 
This anomalous phase is shown to be a minus sign, and hence leads to the $\mathbb{Z}_2$ classification.  
In the second example, by computing the global gravitational anomaly 
\cite{Alvarez1983, Witten1985}
of the Majorana surface states of class D+R$_{+}$ TSCs, 
we study the ``collapse'' of non-interacting classification. 
The resulting projected theories are then shown to be anomalous under large diffeomorphisms (coordinate transformations).

In a similar vein, 
in Ref.\ \onlinecite{Park2015},  
the (3+1)d Weyl fermion on the surface of
the (4+1)d quantum Hall system is shown to fail to be modular invariant 
in the presence of a background U(1) gauge field.


The rest of the paper is organized as follows.
In the remaining part of this section, we introduce some notations that will be used in the main text.
In Sec.\ \ref{SL(n,Z) invariance for (n-1)+1 D fermion theory}, 
we establish the gauge and diffeomorphism invariance of the (2+1)d Dirac fermion theory defined on a spacetime three-torus,
following Refs.\ \onlinecite{Dolan1998, Dolan2013}.
In Sec.\ \ref{Surface theory of (3+1)d CP symmetric topological insulators}, 
we study (3+1)d TIs protected by electromagnetic $\mathrm{U}(1)$ and CP symmetries. 
The surface theory projected by symmetries is shown to be anomalous,
as its (projected) partition function is not invariant under large $\mathrm{U}(1)$ gauge transformations,
but picks up a minus sign, characterizing the $\mathbb{Z}_2$ classification of the bulk phase.
(3+1)d TSCs protected by reflection symmetry are studied in Sec.\ \ref{Surface theory of (3+1)d reflection symmetric crystalline topological superconductors},
where we discuss the invariance/non-invariance of the surface partition function, defined on the three-torus and its descendants generated by the orientifold projection, 
under large diffeomorphisms.
We then conclude in Sec.\  \ref{Discussion}. 

\subsection*{Notations}
\label{Notations}

The partition functions of the (2+1)d surface theories
discussed in the text 
can be represented in terms of partition functions of (1+1)d theories. 
Here, we summarize the properties of these (1+1)d partition functions.

The partition function 
of a (1+1)d chiral fermion (Weyl fermion) on the two-torus $T^2$  with the modular parameter ${\tau}={\tau_1}+i{\tau_2}\in\mathbb{C}$,
 in the presence of spatial $\mathrm{U}(1)$ flux $a$ and temporal $\mathrm{U}(1)$ flux $b$, 
is defined as
\cite{Hsieh2014a, Mirror}
\begin{align}
A^{R}_{[a, b]}({\tau})
&\equiv
 \frac{1}{\eta({\tau})}
\vartheta
\left[
 \begin{array}{c}
  a - 1/2\\
  b-1/2
 \end{array}
\right]
(0,{\tau}),
\nonumber\\
A^{L}_{[a,b]}({\tau})
&=
\left(A^{R}_{[a, b]}({\tau})\right)^*,
\label{2d_chiral_partition function}
\end{align}
where
$\eta({\tau})$ is the Dedekind eta function and 
$\vartheta
\bigl[\begin{smallmatrix}
\alpha\\ \beta
\end{smallmatrix} \bigr]
(v,\tau)$  
is the theta function with characteristics. 
$A^R_{[a, b]}(\tau)$ has the following properties:
\begin{align}
&A^R_{[a,b]}(\tau)
=A^R_{[a+1,b]}(\tau)
= e^{-2\pi i(a-1/2)}A^R_{[a,b+1]}(\tau),
\nonumber\\
&A^R_{[a,b]}({\tau}+1)
= e^{ - \pi i (a^2 -1/6) } A^R_{[a,b+a]}({\tau}),
\nonumber\\
&A^R_{[a,b]}\left(-1/\tau\right)
= e^{ -2\pi i (-a+1/2)(b-1/2) } A^R_{[-b,a]}({\tau}).
 \label{2d_chiral_partition functions__properties}
\end{align}

The partition function of a (1+1)d massive Dirac fermion on $T^2$
with twisted boundary conditions (fluxes $a$ and $b$) is given by
the ``massive theta function'' 
$\Theta_{[a, b]}({\tau}; m)$:
\cite{Takayanagi2002, Sugawara2003}
\begin{align}
\label{massive_theta_func}
&\Theta_{[a, b]}({\tau}; m)
\equiv
 e^{4\pi \tau_2 \Delta(m;a)} 
 \nonumber\\
 &\quad\times
 \prod_{s\in\mathbb{Z}+a}
 \left| 1 - e^{-2\pi\tau_{2}
 \sqrt{m^2+s^2}+2\pi i\tau_{1}s+2\pi i b } \right|^2,
\end{align}
where $\Delta(m;a)$ is the regularized zero-point energy:

\begin{align}
 &\Delta(m;a)
 \equiv
 \frac{1}{2}\sum_{s\in\mathbb{Z}+a}
 \sqrt{ 
  m^2+s^2
 }
 - \frac{1}{2}\int_{-\infty}^{\infty}dk\sqrt{m^2+k^2} 
 \nonumber\\
&\quad\quad= 
-\frac{1}{2\pi^2}\sum_{n=1}^{\infty}\int_0^{\infty}dt e^{-tn^2-\frac{\pi^2m^2}{t}} 
 \cos(2\pi na_x).
 \label{massive_theta_func_2}
 \end{align}
The massive theta function  $\Theta_{[a, b]}({\tau}; m)$ has the following properties:
 \begin{align}
&\Theta_{[a, b]}({\tau}; m)
=\Theta_{[-a, -b]}({\tau}; m)
=\Theta_{[a+r, b+s]}({\tau}; m),
 \quad r, s\in\mathbb{Z},
 \nonumber\\
 &\Theta_{[a, b]}({\tau}+1; m)
 = \Theta_{[a, b+a]}({\tau}; m),
 \nonumber\\
 &\Theta_{[a, b]}\left(-1/\tau; m|{\tau}|\right)
 =\Theta_{[b, -a]}({\tau}; m),
 \nonumber\\
 &\lim_{m\rightarrow0}\Theta_{[a, b]}({\tau}; m)
 =\left| A^R_{[a, b]}(\tau) \right|^2
 =\left| A^L_{[a, b]}(\tau) \right|^2.
 \label{massive_theta_func_properties}
\end{align}

\section{Large U(1) gauge and diffeomorphism invariance of (2+1)d fermion theory}
\label{SL(n,Z) invariance for (n-1)+1 D fermion theory}

In this section, 
we quantize the (2+1)d free Dirac fermion theory on a flat spacetime three-torus $T^3$,
in the presence of background U(1) gauge field and metric.
The invariance of the partition function under large U(1) gauge transformations and
3d modular transformations $\mathrm{SL}(3,\mathbb{Z})$, the mapping class group of $T^3$, will be established. 
A discussion for the 2d modular invariance of the Dirac fermion theory on  two torus $T^2$, as a warm up, is reviewed in Appendix  \ref{The Dirac fermion theory on two torus $T^2$}.

We closely follow the analysis and notations in Ref.\ \onlinecite{Dolan1998}. 
(See also Refs.\ \onlinecite{Dolan2013} for related works.)
In Ref.\ \onlinecite{Dolan1998},
the partition function of a chiral self-dual two-form gauge field on a 6d spacetime torus $T^6$,  
and its invariance under $\mathrm{SL}(6,\mathbb{Z})$, the mapping class group of the six-torus,  
was studied.  
In Ref.\ \onlinecite{Dolan1998}, the theory is quantized (regularized) in a way manifestly symmetric 
under $\mathrm{SL}(5,\mathbb{Z})$.   
It was then shown that the partition function has an additional 
$\mathrm{SL}(2, \mathbb{Z})$ invariance,
and together with the $\mathrm{SL}(5, \mathbb{Z})$ invariance, 
the full $\mathrm{SL}(6, \mathbb{Z})$ invariance was proven.
By properly adopting this strategy,  
we show the $\mathrm{SL}(3,\mathbb{Z})$ invariance 
and the large gauge invariance of the (2+1)d Dirac fermion theory. 

While our focus in this section is on the complex or Dirac fermion, 
the case for real or Majorana fermions can be studied in a similar way. 
The modular properties studied here are expected to be straightforwardly generalized
to higher dimensions, e.g., $\mathrm{SL}(n,\mathbb{Z})$ invariance for the partition function.

\subsection*{The Dirac fermion theory on three torus $T^3$}

\subsubsection{Background metric}

A flat three-torus is parameterized by five ``modular parameters'', 
$R_{1,2}/R_0$, $\alpha$, $\beta$, and $\gamma$, 
where $R_{\mu}$ are the radii for the $\mu$-th directions,
and $\alpha,\beta,\gamma$ and  related to the angles between directions $0$ and $1$,  $1$ and $2$, and $0$ and $2$, respectively. 
The dreibein is given by ($\mu, A=0,1,2$)
\begin{align}
 {e^A}_{\mu} 
 &=
 \left(
 \begin{array}{ccc}
 R_0 & 0 & 0 \\
 0 & R_1 & 0 \\
 0 & 0 & R_2 
 \end{array}
 \right)
 \left(
  \begin{array}{ccc}
 1 & 0 & 0 \\
 -\alpha & 1 & 0 \\
 -\gamma & -\beta & 1 
 \end{array}
 \right)
 \nonumber\\
 &=
 \left(
  \begin{array}{ccc}
 R_0 & 0 & 0 \\
 -\alpha R_1 & R_1 & 0 \\
 -\gamma R_2 & -\beta R_2 & R_2 
 \end{array}
 \right),
\end{align}
and its inverse is given by
\begin{align}
 {e^{\star}_A}^{\mu}=
  \left(
 \begin{array}{ccc}
 \frac{1}{R_0} & \frac{\alpha}{R_0} & \frac{\alpha\beta+\gamma}{R_0} \\
 0 & \frac{1}{R_1} & \frac{\beta}{R_1} \\
 0 & 0 & \frac{1}{R_2} 
 \end{array}
 \right),
\end{align}
such that ${e^A}_{\mu}{e^{\star}_A}^{\nu}={\delta_{\mu}}^{\nu}$ and ${e^A}_{\mu}{e^{\star}_B}^{\mu}={\delta^A}_B$. 
The Euclidean metric is 
\begin{align}
 g_{\mu\nu} 
 &=
 {e^A}_{\mu}{e^B}_{\nu}\delta_{AB}
 \nonumber\\
 &=
  \left(
\begin{array}{ccc}
 R_0^2+\alpha^2R_1^2+\gamma^2R_2^2 & -\alpha R_1^2+\beta\gamma R_2^2 & -\gamma R_2^2 \\
 -\alpha R_1^2+\beta\gamma R_2^2 & R_1^2+\beta^2R_2^2 & -\beta R_2^2 \\ 
 -\gamma R_2^2 & -\beta R_2^2 & R_2^2 
 \end{array}
 \right),
\end{align}
and the line element is given by $ds^2=g_{\mu\nu}d\theta^{\mu}d\theta^{\nu}$,
where $0\leq \theta^{\mu}\leq 2\pi$ are angular variables.

The group $\mathrm{SL}(3,\mathbb{Z})$ is generated by two modular transformations:
\cite{Generators}
\begin{align}
\label{U1 and U2}
 U_1 =
 \left(
 \begin{array}{ccc}
 0 & 0 & 1 \\
 1 & 0 & 0 \\
 0 & 1 & 0 
 \end{array}
\right),
\quad
 U_2 =
 \left(
 \begin{array}{ccc}
 1 & 1 & 0 \\
 0 & 1 & 0 \\
 0 & 0 & 1 
 \end{array}
\right).
\end{align}
The dreiben and metric are transformed as  
\begin{align}
 {e^A}_{\mu} 
 &\overset{L}{\longrightarrow} 
 \ {{( e L^T)}^A}_{\mu}
 = {L_{\mu}}^{\rho} {e^A}_{\rho},
 \nonumber\\
 {e^{\star}_A}^{\mu}
 &\overset{L}{\longrightarrow} 
 {{(e^{\star}L^{-1})}_A}^{\mu}
 =  {e^{\star}_A}^{\rho}{{(L^{-1})}_{\rho}}^{\mu},
 \nonumber\\
 g_{\mu\nu}  
 &\overset{L}{\longrightarrow} 
 \ {( L g L^T)}_{\mu\nu}
 = {L_{\mu}}^{\rho} {L_{\nu}}^{\sigma}  g_{\rho\sigma},
\end{align}
for any $\mathrm{SL}(3,\mathbb{Z})$ elements $ L = U^{n_1}_1 U^{n_2}_2 U^{n_3}_1\cdots$.
In particular,
\begin{widetext}
\begin{align}
 g_{\mu\nu}  
 &\overset{U_2}{\longrightarrow} 
 \  {( U_2 g U^T_2)}_{\mu\nu} 
 =
 \left(
  \begin{array}{ccc}
 R_0^2+(\alpha-1)^2R_1^2+(\gamma+\beta)^2R_2^2 & -(\alpha-1) R_1^2+\beta(\gamma+\beta) R_2^2 & -(\gamma+\beta) R_2^2 \\
 -(\alpha-1) R_1^2+\beta(\gamma+\beta) R_2^2 & R_1^2+\beta^2R_2^2 & -\beta R_2^2 \\ 
 -(\gamma+\beta) R_2^2 & -\beta R_2^2 & R_2^2 
 \end{array}
 \right),
\end{align}
which corresponds to the changes
\begin{align}
\label{mod_under_U_2}
\alpha \rightarrow \alpha-1, \quad
\gamma \rightarrow \gamma+\beta 
&
\quad 
\text{(while $R_0$, $R_1$, $R_2$, and $\beta$ are unchanged)}.
\end{align}
The less trivial generator $U_1$ can be further decomposed into two transformations as
\begin{align}
\label{U1' and M}
 U_1 
 &= 
 U'_1M,
 \quad 
 U'_1
=
 \left(
 \begin{array}{ccc}
 0 & -1 & 0 \\
 1 & 0 & 0 \\
 0 & 0 & 1 
 \end{array}
\right),
\quad
 M =
 \left(
 \begin{array}{ccc}
 1 & 0 & 0 \\
 0 & 0 & -1 \\
 0 & 1 & 0 
 \end{array}
\right).
\end{align}
The transformation $U^{\prime}_1$ acts on the metric as
\begin{align}
 g_{\mu\nu}  
 &\overset{U'_1}{\longrightarrow} 
 \  {( U'_1 g U'^T_1)}_{\mu\nu} 
 =
 \left(
  \begin{array}{ccc}
 R_1^2+\beta^2R_2^2 & \alpha R_1^2-\beta\gamma R_2^2 & \beta R_2^2 \\
 \alpha R_1^2-\beta\gamma R_2^2 & R_0^2+\alpha^2R_1^2+\gamma^2R_2^2 & -\gamma R_2^2 \\ 
 \beta R_2^2 & -\gamma R_2^2 & R_2^2 
 \end{array}
 \right),
\end{align}
which corresponds to the changes
\begin{align}
\label{mod_under_U_1'}
&R_0 \rightarrow R_0/|{\tau_{2d}}|, \quad
R_1 \rightarrow R_1|{\tau_{2d}}|, \quad
\alpha \rightarrow -\alpha/|{\tau_{2d}}|^2 \quad \text{(or ${\tau_{2d}} \rightarrow -1/\tau_{2d}$)}, 
\quad 
\gamma \rightarrow -\beta, \quad
\beta \rightarrow \gamma \quad
\text{(while $R_2$ is unchanged)},
\end{align}
with
\begin{align}
{\tau_{2d}}\equiv\alpha+ir_{01},
\quad
 r_{\mu\nu}\equiv R_{\mu}/R_{\nu},
\end{align} 
while $M$ acts on the metric as 
\begin{align}
 g_{\mu\nu}  
 &\overset{M}{\longrightarrow} 
 \  {( M g M^T)}_{\mu\nu}
=
 \left(
  \begin{array}{ccc}
 R_0^2+\alpha^2R_1^2+\gamma^2R_2^2 & \gamma R_2^2 & -\alpha R_1^2+\beta\gamma R_2^2 \\
 \gamma R_2^2 & R_2^2 & \beta R_2^2 \\ 
 -\alpha R_1^2+\beta\gamma R_2^2 & \beta R_2^2 & R_1^2+\beta^2R_2^2
 \end{array} 
 \right).
\end{align}

The Euclidean action for the Dirac fermion on $T^3$ is then given by
\begin{align}
  S_E
 &=  \frac{1}{(2\pi)^2}\int d^3 \theta\, \left(\det{e}\right) \bar{\psi} \left(\Gamma^A {e^{\star}_A}^{\mu} \frac{\partial}{\partial\theta^{\mu}} \right) \psi \nonumber\\
 &= \frac{1}{(2\pi)^2}
 \int_0^{2\pi R_0} d{\tau} \int_0^{2\pi R_1} dx \int_0^{2\pi R_2} dy\, 
 \bar{\psi} 
 \left[
 \Gamma^0\partial_{{\tau}}+\alpha\frac{R_1}{R_0}\Gamma^0\partial_{x}+(\alpha\beta+\gamma)\frac{R_2}{R_0}\Gamma^0\partial_{y}
 +\Gamma^1\partial_{x}+\beta\frac{R_2}{R_1}\Gamma^1\partial_{y}+\Gamma^2\partial_{y}
 \right] \psi,
 \end{align}
\end{widetext}
 where 
$\psi$ is the two-component Dirac field,
$ \bar{\psi}=\psi^{\dag}\Gamma^0$,
 ${\tau}=R_0\theta^0$, $x=R_1\theta^1$, $y=R_2\theta^2$,
 and  the gamma matrices $\Gamma^A$ satisfy $\{\Gamma^A, \Gamma^B\}= 2\delta^{AB}$.

\subsubsection{Background flux}

In addition to the background metric, 
we also introduce the background U(1) gauge field (flux) on $T^3$
to twist the boundary conditions of the Dirac fermion theory.
More specifically, in the path integral language, 
we consider the boundary conditions 
\begin{align}
\label{flux_twisted_BCs}
& \psi({\tau},x+2\pi R_1,y) = e^{ 2\pi i a_x}\psi({\tau},x,y),
\nonumber \\
& \psi({\tau},x,y+2\pi R_2) = e^{ 2\pi i a_y}\psi({\tau},x,y),
 \nonumber \\
& \psi({\tau}+2\pi R_0, x- 2\pi \alpha R_1, y-2\pi (\alpha\beta+\gamma)R_2) 
\nonumber \\
&\qquad 
= e^{ 2\pi i a_{\tau}}\psi(\tau,x,y), 
\end{align}
where $(a_{\tau},a_x,a_y)\equiv \mathsf{a}$ represents the background U(1) gauge field twisting the boundary conditions.

\subsubsection{Partition function} 

We now quantize the (2+1)d theory and compute the properly regularized partition function, 
denoted by 
$Z_{[\mathsf{a}]}(g)$, 
which depends on the background flux $\mathsf{a}$ and metric $g$. 
The partition function can be evaluated by the path integral on $T^3$, 
$
 Z_{[\mathsf{a}]}(g)=\int \mathcal{D}[\psi^{\dag},\psi]  \exp(-S_E)
$,
with $\psi$ satisfying the twisted boundary conditions (\ref{flux_twisted_BCs}),
or alternatively, in the operator language,  by the trace
\begin{align}
\label{part_func}
 Z_{[\mathsf{a}]}(g)
 &=
 \mathrm{Tr}_{a_xa_y} \left[
 e^{2\pi i (a_{\tau}-1/2) F} 
 e^{-2\pi R_0H' }
 \right],
\end{align}
where $H'$ is the "boosted" Hamiltonian (in the presence of non-vanishing angles $\alpha$, $\beta$, and $\gamma$) obtained from $S_E$
 and given by
 \begin{align} 
 H'=  H - i\alpha\frac{R_1}{R_0} P_x- i(\alpha\beta+\gamma)\frac{R_2}{R_0} P_y,
 \label{boosted_H'_T^3}
 \end{align}
with
\begin{align}
\label{H&P_i}
 H &=
\frac{1}{(2\pi)^2} 
 \int dxdy\, 
 \bar{\psi} \left( 
 \Gamma^1\partial_{x}+\beta\frac{R_2}{R_1}\Gamma^1\partial_{y}+\Gamma^2\partial_{y}
 \right) \psi,
 \nonumber\\
 P_i &=
 \frac{1}{(2\pi)^2} 
 \int dxdy\, 
 \psi^{\dagger} (-i\partial_i\psi),
 \quad
 i=x, y,
\end{align}
being the Hamiltonian and momenta.
$\mathrm{Tr}_{a_x, a_y}$ means the trace is taken over the
Fock space of the fermion theory for the
spatial boundary conditions specified by $a_x$ and $a_y$.
The twisted boundary condition in the ${\tau}$-direction is implemented by an operator insertion
$\exp [2\pi i (a_{\tau}-1/2)F]$, where $F$ is the fermion number operator.

The fermion field operator satisfies the canonical anticommutation relation 
\begin{align}
& \{\psi_{\alpha}(\mathsf{r}), \psi_{\beta}^{\dag}(\mathsf{r}') \}
 =
 (2\pi)^2 \delta_{\alpha\beta}
 \sum_{m_1, m_2\in \mathbb{Z}}
 \nonumber \\
 &\quad 
 \times 
 \delta(x-x'+2\pi m_1 R_1)
 \delta(y-y'+2\pi m_2 R_2),
\end{align}
where $\mathsf{r}=(x,y)$ and $\alpha, \beta$ are spinor indices.
The trace can be evaluated explicitly by the Fourier mode expansion of the fermion field operator.
With the twisted boundary conditions, 
the fermion field operator is expanded as
\begin{align}
 \psi(\mathsf{r}) &= 
 \frac{1}{\sqrt{R_1 R_2}}
 \sum_{s_x\in \mathbb{Z}+a_x} 
 \sum_{s_y \in \mathbb{Z}+ a_y}
 e^{  i x \frac{ s_x}{R_1}+ 
   i y \frac{ s_y}{R_2}} 
 \tilde{\psi}(\mathsf{s}),  
 \end{align}
where $\mathsf{s}=(s_x,s_y)$ and
\begin{align}
\{ \tilde{\psi}_{\alpha}(\mathsf{s}), \tilde{\psi}_{\beta}^{\dag}(\mathsf{s}') \}
=
\delta_{\alpha\beta}\delta_{\mathsf{s}\mathsf{s}'}. 
\end{align}
Correspondingly, the Hamiltonian can be expanded as 
\begin{align}
 &
 H= 
 \sum_{s_x\in \mathbb{Z}+a_x} \sum_{s_y \in \mathbb{Z}+ a_y}
 \tilde{\psi}^{\dag}(\mathsf{s})
 \mathcal{H}(\mathsf{s})
 \tilde{\psi}(\mathsf{s}), 
 \nonumber \\
 &
 \mathcal{H}(\mathsf{s})
 =
 \Gamma^0
 \left[
 \Gamma^1 
 \frac{is_x}{R_1}
 +
 \beta 
 \frac{R_2}{R_1}
 \Gamma^1  \frac{is_y}{R_2}
 +
 \Gamma^2 
 \frac{is_y}{R_2}
 \right].
\end{align}

The single-particle Hamiltonian $\mathcal{H}(\mathsf{s})$ can be diagonalized
with eigenvectors $\vec{u}_{\pm}(\mathsf{s})$ and
eigenvalues $\pm \varepsilon(\mathsf{s})$:
 \begin{align}
 & 
 \mathcal{H}(\mathsf{s}) \vec{u}_{\pm} (\mathsf{s})  
 = \pm \varepsilon(\mathsf{s}) \vec{u}_{\pm} (\mathsf{s}),
 \nonumber \\
 &
\varepsilon(\mathsf{s}) 
=
 \sqrt{g_2^{ij}s_is_j}
= \sqrt{\left(\frac{s_x}{R_1}+\beta\frac{s_y}{R_1}\right)^2+\left(\frac{s_y}{R_2}\right)^2 },
\end{align}
where
\begin{align}
\label{g2_inverse}
 g_2^{ij}
 \equiv
\left(
 \begin{array}{cc}
  g_{11} & g_{12} \\
  g_{21} & g_{22} \\
 \end{array}
 \right)^{-1}
 =\left(
 \begin{array}{cc}
  \frac{1}{R_1^2} & \frac{\beta}{R_1^2} \\
  \frac{\beta}{R_1^2} & \frac{\beta^2}{R_1^2}+\frac{1}{R_2^2} \\
 \end{array}
 \right).
\end{align}
The Hamiltonian can be diagonalized by the eigen basis
$\chi(\mathsf{s}):=[\chi_+(\mathsf{s}), \chi_-(\mathsf{s})]^T$, 
which are related to the original fermion operators $\tilde{\psi}(\mathsf{s})$ as 
\begin{align}
 \left[
 \begin{array}{c}
 \tilde{\psi}_{1}(\mathsf{s}) \\
 \tilde{\psi}_{2}(\mathsf{s}) 
 \end{array}
 \right]
 =
 \left[
 \begin{array}{cc}
 u^{\ }_{1+}(\mathsf{s})& u^{\ }_{1-}(\mathsf{s}) \\
 u^{\ }_{2+}(\mathsf{s})& u^{\ }_{2-}(\mathsf{s}) \\
 \end{array} 
 \right]
 \left[
 \begin{array}{c}
 \chi_{+}(\mathsf{s}) \\
 \chi_{-}(\mathsf{s})
 \end{array} 
 \right],
 \nonumber \\
 \left[
 \begin{array}{c}
 \chi_{+}(\mathsf{s}) \\
 \chi_{-}(\mathsf{s}) 
 \end{array}
 \right]
 =
 \left[
 \begin{array}{cc}
 u^{*}_{1+}(\mathsf{s})& u^{*}_{2+}(\mathsf{s}) \\
 u^{*}_{1-}(\mathsf{s})& u^{*}_{2-}(\mathsf{s}) \\
 \end{array} 
 \right]
 \left[
 \begin{array}{c}
 \tilde{\psi}_{1}(\mathsf{s}) \\
 \tilde{\psi}_{2}(\mathsf{s})
 \end{array} 
 \right]. 
 \label{fermion_eigenbasis}
\end{align}
The Hamiltonian in the eigen basis is given by
\begin{align}
 H &= \sum_{\mathsf{s}}
 \varepsilon(\mathsf{s})
 \left[
 \chi^{\dag}_{+}(\mathsf{s}) \chi^{\ }_{+}(\mathsf{s})
-
 \chi^{\dag}_{-} (\mathsf{s})\chi^{\ }_{-}(\mathsf{s})
\right] 
\nonumber \\
  &=
 \sum_\mathsf{s} \varepsilon(\mathsf{s})
 \left[
 \chi^{\dag}_{+}(\mathsf{s}) \chi^{\ }_{+}(\mathsf{s})
 +
\chi^{\ }_{-}(x) \chi^{\dag}_{-}(\mathsf{s})
 \right]
- 
 \sum_\mathsf{s} \varepsilon(\mathsf{s}) 
 \nonumber \\
 & =
 :H: + E_{\mathrm{GS}} 
\end{align}
where $:\cdots:$ is the normal ordering with respect to the Fock vacuum
and $E_{\mathrm{GS}}= -\sum_{\mathsf{s}}\varepsilon (\mathsf{s})$ is the ground-state energy. 

The ground-state energy needs to be properly regularized.
As shown in Appendix\ \ref{Regularization of the ground-state energy}, we have
\begin{align}
E_{\mathrm{GS}[\mathsf{a}]}(g)
&=
-\sum_\mathsf{s}\sqrt{g_2^{ij}s_is_j}
\nonumber\\
&=
 \frac{1}{4\pi^2}\sqrt{\det(g_{2 ij})} 
 \sum_{\mathsf{n}\neq 0\in\mathbb{Z}^2} 
 \frac{\cos\left(2\pi ia^i n_i\right)}{\left(g_2^{ij}n_in_j \right)^{\frac{3}{2}}},
 \label{GS regularized}
\end{align}
where 
$s_{x,y}$ are separated into their integral and fractional parts as
\begin{align}
s_{i}= n_{i} + a_{i},
\quad
n_{i}\in \mathbb{Z}, 
\quad 
i=x,y. 
\end{align}
Similarly, the ground-state momentum and the fermion number can be regularized as 
 \begin{align}
 (P_i)_\mathrm{GS}
&= 
\sum_{s_i} s_i 
= \sum_{s_i>0} s_i +  \sum_{s_i<0} s_i
\nonumber \\
&= \zeta(-1, a_i) - \zeta(-1, 1-a_i)
=0,
\nonumber\\
F_\mathrm{GS}
&= 
\sum_\mathsf{s}1-\sum_\mathsf{s}1
=0, 
\label{g_2^{ij}}
\end{align}
where $\zeta(s, x)=\sum^{\infty}_{n=0}(n+x)^{-s}$ is the Hurwitz zeta function defined by
analytic continuation from the region $\mathrm{Re} (s) > 1$ and we have used
$\zeta(-1, x)=\frac{1}{24}-\frac{1}{2}\left(x-\frac{1}{2}\right)^2$.

With the regularization, 
the partition function (with boundary conditions twisted by $a_x$ and $a_y$), given by the trace in  
(\ref{part_func}), is evaluated as
 \begin{align}
& 
Z_{[\mathsf{a}]}(g)
=e^{ -2\pi R_0E_\mathrm{GS}
 }
 \prod_{s_y\in\mathbb{Z}+a_y} \prod_{s_x\in\mathbb{Z}+a_x} 
 \nonumber\\
 &\quad
 \times  
 \left| 
 1 - e^{-2\pi R_0 \varepsilon(\mathsf{s})+2\pi i\alpha s_x+2\pi i(\alpha\beta+\gamma)s_y+2\pi ia_{\tau}}
 \right|^2,
 \label{full part fn}
 \end{align}
which can also be expressed as the infinite product of massive theta functions defined in (\ref{massive_theta_func}):
\begin{align}
\label{full part fn_2}
&Z_{[\mathsf{a}]}(g)
= 
 \prod_{s_{y}\in\mathbb{Z}+a_y} 
  \Theta_{[a_{x}+\beta s_y, a_{\tau}+\gamma s_y]} 
 \left(
\tau_{2d}; r_{12} s_y \right).
\end{align}

We now demonstrate the invariance of the partition function
under large U(1) gauge transformations and modular transformations. 

 
\subsubsection{Large U(1) gauge invariance of the partition function}
\label{Large U(1) gauge invariance of the partition function}

We first check the invariance of the partition function 
under large U(1) gauge transformations $a_{x,y,{\tau}}\to a_{x,y,{\tau}}+1$.
The invariance under $a_{x, \tau}\to a_{x, \tau}+1$ is obvious from Eq.\ (\ref{full part fn_2}),
using the properties of the massive theta function listed in
(\ref{massive_theta_func_properties}). 
To check the invariance of the partition function under 
$a_{y}\to a_{y}+1$,
we note that this amounts to a simple shift $s_{y}\to s_{y}+1$ in the infinite product in 
Eq.\ (\ref{full part fn_2}).
To sum up, we conclude the large U(1) gauge invariance of the partition function.

\subsubsection{Modular invariance of the partition function}

By using the expressions 
(\ref{full part fn}) and (\ref{full part fn_2})
of the partition function,
we can show that $Z_{[\mathsf{a}]}(g)$ has the following property:
\begin{align}
 Z_{[L\mathsf{a}]}(LgL^T)
 &=
 Z_{[\mathsf{a}]}(g),
 \nonumber\\ 
 \text{or}\quad
 Z_{[\mathsf{a}]}(LgL^T)
 &=
 Z_{[L^{-1}\mathsf{a}]}(g),
 \quad
 \label{SL(3,Z)_invariance}
\end{align}
where 
$L\in \mathrm{SL}(3,\mathbb{Z})$.
This means the Dirac fermion, when coupled to both background U(1) gauge field and metric, is anomaly-free under any large diffeomorphisms (together with the induced gauge transformations) on $T^3$.
The claim (\ref{SL(3,Z)_invariance}) can be shown 
by checking how $Z_{[\mathsf{a}]}(g)$ transforms under $U_1=U'_1M$ and $U_2$, defined in Eqs.\ (\ref{U1 and U2}) and (\ref{U1' and M}).
Here we leave the detail of the derivation to Appendix\ 
\ref{Derivation of the claim (ref{SL(3,Z)_invariance})}.

We now show that the partition function, once projected by the fermion number parity only [i.e., in the absence of U(1) gauge field], 
is modular invariant.
We consider the sum over all periodic/antiperiodic boundary conditions, 
which are specified by boundary conditions 
$\mathsf{a}=(0, 0, 0)$, $(\frac{1}{2}, 0, 0)$, $\ldots$, 
$(\frac{1}{2}, \frac{1}{2}, \frac{1}{2})$,
corresponding to the $2^3=8$ spin structures when defining fermions (spinors) on $T^3$.
The resulting total partition function is given by
\begin{align}
\mathcal{Z}^{tot}(g)=\sum_{a_{x, y, \tau} = 0, 1/2} \epsilon_{\mathsf{a}} Z_{[\mathsf{a}]}(g),
\end{align}
where $\epsilon_{\mathsf{a}}$ are weights (``discrete torsion'') assigned to different sectors 
with partition functions twisted by $\mathsf{a}$. 
From Eq.\ (\ref{SL(3,Z)_invariance}), 
we see that
by choosing $\epsilon_{\mathsf{a}}=1$ for all $\mathsf{a}$,  
the total partition function is 
modular invariant:
\begin{align}
\mathcal{Z}^{tot}(LgL^T)=\mathcal{Z}^{tot}(g),\ L\in \mathrm{SL}(3,\mathbb{Z}).
\end{align}


\section{Surface theory of (3+1)d CP symmetric topological insulators} 
\label{Surface theory of (3+1)d CP symmetric topological insulators}

Based on the result from the previous section, now we can compute quantum anomalies of an anomalous surface theory and interpret them as a signal of the existence of the nontrivial bulk SPT phases.
In this section,  
we identify a global $\mathrm{U}(1)$ gauge anomaly
of the surface theory of (3+1)d CP (charge conjugation $\times$ reflection) symmetric TIs,
which are related to, by CPT-theorem, (3+1)d time-reversal symmetric TIs.

\subsection{Surface theory}
Let us consider the following surface Dirac Hamiltonian of (3+1)d TIs:
\begin{align}
 H =
 \frac{1}{(2\pi)^2}
 \int dx dy\, \psi^{\dag}(\mathsf{r}) (-i \sigma_2 \partial_x- i \sigma_1 \partial_y)  \psi(\mathsf{r}), 
 \label{surface Dirac cp}
\end{align}
where $\sigma_{1,2,3}$ are the Pauli matrices,
the spatial coordinate $\mathsf{r}=(x,y)\in [0, 2\pi R_1)\times[0,2\pi R_2)$ parameterizes the 2d surface, and 
$\psi(\mathsf{r})=[\psi_1(\mathsf{r}), \psi_2(\mathsf{r})]^T$ 
and 
$\psi^{\dag}(\mathsf{r})= [\psi^{\dag}_1(\mathsf{r}), \psi^{\dag}_2(\mathsf{r})]$ 
are two-component fermion annihilation and creation operators, respectively.  
The Hamiltonian is invariant under the following time-reversal symmetry:
\begin{align}
 \mathscr{T} \psi (\mathsf{r})\mathscr{T}^{-1} = i \sigma_2 \psi(\mathsf{r}), 
 \quad
 \mathscr{T}^2 = (-1)^F,
\end{align}
where $F$ is the fermion number operator.
This time-reversal symmetry prohibits the mass term $\psi^{\dag}\sigma_3 \psi$ since 
$ \mathscr{T} 
 \psi^{\dag}\sigma_3 \psi 
 \mathscr{T}^{-1}
=
-
 \psi^{\dag}\sigma_3 \psi 
$.

Alternatively,
in the following, we will take the Dirac Hamiltonian  (\ref{surface Dirac cp})
as a surface theory of bulk CP symmetric TIs. 
By CPT-theorem, 
the classification of CP symmetric TIs are
expected to be the same as
the classification of time-reversal symmetric TIs.
That is, 
CP symmetric insulators in (3+1)d are also classified by a $\mathbb{Z}_2$ topological number. 
\cite{Hsieh2014b}
Within the surface theory, the action of CP symmetry is given by
\begin{align}
&
(\mathscr{CP}) \psi(x,y) (\mathscr{CP})^{-1} = \sigma_3 [\psi^{\dag}(x, 2\pi R_2-y)]^T,
\nonumber \\
&
(\mathscr{CP}) \psi^{\dag}(x,y)(\mathscr{CP})^{-1}= \psi(x, 2\pi R_2-y)^T \sigma_3,
 \nonumber \\
& 
 (\mathscr{C}\mathscr{P})^2=1. 
\end{align}
[This is the only CP symmetry of the Dirac kinetic term 
$\mathcal{H}(k_x, k_y) = k_x \sigma_2 + k_y \sigma_1$
since $\sigma_3\mathcal{H}^T(-k_x, k_y)\sigma_3^{-1} = -\mathcal{H}(k_x, k_y)$.]
Fermion bilinears in the surface theory are transformed as
$
 \psi^{\dag} M \psi
 \to 
 \psi^T U^{\dag} M U (\psi^{\dag})^T
 =
 - \psi^{\dag} U^T M^T U^* \psi.
$
In particular, the mass is odd under CP;
the surface theory, at least at quadratic level, 
cannot be gapped without breaking symmetries,
$\mathrm{U}(1)\rtimes\mathrm{CP}$. 
On the other hand, a CP preserving mass exists if we double this theory (or more generally if the number of the Dirac fermions is even),
and the corresponding surface theory can be gapped.
Since massive fermions can always be regularized to construct a well-defined quantum theory, an even number of the surface fermions (\ref{surface Dirac cp}) is always anomaly-free (while preserving the symmetries).
However, an odd number of the surface fermions may suffer from anomalies.
In the following, we will identify a quantum anomaly 
of the surface theory (\ref{surface Dirac cp}) under large $\mathrm{U}(1)$ gauge transformation  
when $\mathrm{CP}$ symmetry is strictly enforced.

\subsection{Projected partition function by CP symmetry}

We now consider CP projection of the surface theory 
(\ref{surface Dirac cp}) 
and ask if the
projected theory is still invariant under large gauge transformations (as we have seen in the theory without symmetry projection).
This leads to formulating the fermion theory on unorientable spacetime manifolds such as $S^1 \times K$, where $K$ is the Klein bottle. 
As our main focus here is on large gauge transformations but not on modular transformations,
the parameters $\alpha$, $\beta$, and $\gamma$ are set to zero in the following discussion. 
Also, the twisted boundary conditions by U(1) are consistent with CP symmetry
only when $a_{\tau, x} = 0, 1/2$,
while $a_y$ is not constrained by CP symmetry.
We will thus study the large gauge transformation
$a_y\to a_y+1$. 

The partition function of the surface theory projected by CP is given by the trace
\begin{align}
\label{trace_CP_proj}
\mathrm{Tr}_{a_x,a_y}\
 \left[\frac{1}{2}(1+ \mathscr{CP}) e^{2\pi i (a_{\tau}-1/2) F}
 e^{-2\pi R_0 H} \right].
\end{align}
Upon projection by CP, 
we will focus on the CP symmetric boundary conditions
$a_{\tau, x} = 0, 1/2$ and $a_y \in [0,1)$.
The first term in the projected trace, which is invariant under $a_{y}\to a_{y}+1$,  is already discussed in Sec.\ \ref{Large U(1) gauge invariance of the partition function}.  
Our focus below will be the second term in the projected trace,
which we call the CP twisted partition function: 
\begin{align}
\label{trace_CP_twisted}
 Z^{\mathrm{CP}}_{[\mathsf{a}]} 
 &= 
 \mathrm{Tr}_{a_x,a_y}\, 
 \mathscr{CP} e^{2\pi i (a_{\tau}-1/2) F}
 e^{-2\pi R_0 H}.
\end{align}

 Because of CP symmetry, from the eigenvectors $\vec{u}_{\pm}$ at $\mathsf{s}$, 
we can construct eigenvectors at $\bar{\mathsf{s}}= (-s_x, s_y)$:
\begin{align}
 \mathcal{H}(\bar{\mathsf{s}}) \sigma_3 \vec{u}^*_{\pm} (\mathsf{s}) & 
 = \mp \varepsilon(\mathsf{s}) \sigma_3 \vec{u}^*_{\pm} (\mathsf{s}).
\end{align}
The CP operator acts on the Fourier components of the original fermion operators as 
\begin{align}
(\mathscr{CP})
\tilde{\psi}(\mathsf{s})
(\mathscr{CP})^{-1}
=
\sigma_3
\left[ \tilde{\psi}^{\dag}(\bar{\mathsf{s}}) \right]^T. 
\end{align}
On the other hand, the CP action on the eigen basis $\chi^{\dag}_{\pm}, \chi^{\ }_{\pm}$ is deduced as 
\begin{align}
&\quad
(\mathscr{CP})
\chi(\mathsf{s})
(\mathscr{CP})^{-1}
\nonumber \\
& 
=
\left[
\begin{array}{cc}
\langle {u}_{+}(\mathsf{s})|\sigma_3 K |{u}_+(\bar{\mathsf{s}})\rangle & 
\langle {u}_{+}(\mathsf{s})|\sigma_3 K|{u}_-(\bar{\mathsf{s}})\rangle  \\
\langle {u}_{-}(\mathsf{s})|\sigma_3 K|{u}_+(\bar{\mathsf{s}})\rangle  
&
\langle {u}_{-}(\mathsf{s})|\sigma_3 K|{u}_-(\bar{\mathsf{s}})\rangle   \\
\end{array}
\right]
\left[ \chi^{\dag}(\bar{\mathsf{s}}) \right]^T, 
\end{align}
where 
$K$ is the complex conjugation operator, and 
$\langle {u}_{\pm}(\mathsf{s})|\sigma_3 K |{u}_{\pm}(\bar{\mathsf{s}})\rangle 
= \vec{u}^*_{\pm}(\mathsf{s}) \cdot \sigma_3 \vec{u}^*_{\pm}(\bar{\mathsf{s}})$,
etc.   
Since $\vec{u}_{\pm}(\mathsf{s})$ and $\sigma_3 \vec{u}^*_{\pm}(\bar{\mathsf{s}})$
are both eigenvectors of Hamiltonian $\mathcal{H}(\mathsf{s})$
but with different energies, their overlap should be zero.
Therefore,
\begin{align}
&
( \mathscr{CP}) \chi_+(\mathsf{s}) ( \mathscr{CP})^{-1} = \langle {u}_{+}(\mathsf{s})|\sigma_3 K|{u}_-(\bar{\mathsf{s}})\rangle  \chi^{\dag}_{-}(\bar{\mathsf{s}}),
\nonumber \\
&
( \mathscr{CP}) \chi_-(\mathsf{s}) ( \mathscr{CP})^{-1} 
= \langle {u}_{-}(\mathsf{s})|\sigma_3 K|{u}_+(\bar{\mathsf{s}})\rangle  \chi^{\dag}_{+}(\bar{\mathsf{s}}). 
\end{align}
The transformation law of $\chi_{\pm}$ under CP
depends on the choice of eigen functions $\vec{u}_{\pm}$.  
A choice for the eigenvectors is 
\begin{align}
  |u_{\pm}(\mathsf{s}) \rangle
  &=
  \frac{1}{\sqrt{2}}
  \left[
    \begin{array}{c}
      \pm \Delta^*/|\Delta| \\
      1
    \end{array}
  \right],
  \quad 
\Delta=s_y+{i}s_x. 
\end{align}
For this choice of eigenfunctions, 
\begin{align}
 \langle u_{+}(\mathsf{s}) |\sigma_3 K|u_- (\bar{\mathsf{s}})\rangle 
 &=
 \langle u_{-}(\mathsf{s}) |\sigma_3 K|u_+ (\bar{\mathsf{s}})\rangle 
 =-1.
\end{align}
As we can choose the phase of the eigenvectors freely, 
\begin{align}
    |u_{\pm} (\mathsf{s})\rangle
    &=
    \frac{1}{\sqrt{2}}
    \left[
      \begin{array}{c}
        1 \\
        \pm \Delta/|\Delta|
      \end{array}
    \right]
\end{align}
is also an eigenfunction.   
In this gauge, 
\begin{align}
 \langle u_{+}(\mathsf{s}) |\sigma_3 K|u_- (\bar{\mathsf{s}})\rangle 
 &=
 \langle u_{-}(\mathsf{s}) |\sigma_3 K|u_+ (\bar{\mathsf{s}})\rangle 
 =1.
\end{align}
In either choice, the result can be summarized as
\begin{align}
&
 (\mathscr{CP}) \chi^{\ }_{+}(\mathsf{s}) (\mathscr{CP})^{-1} = 
 \eta_+ \chi^{\dag}_-(\bar{\mathsf{s}}),
 \nonumber \\
 &
 (\mathscr{CP}) \chi^{\ }_{-}(\mathsf{s}) (\mathscr{CP})^{-1} 
 = \eta_- \chi^{\dag}_+(\bar{\mathsf{s}}),
 \label{chi_pm_under_CP}
\end{align}
where $\eta_{\pm}$ is $s$-independent.  
The product $\eta :=\eta_+ \eta_-=1$ 
is gauge invariant.

The CP twisted partition function 
$Z^{\mathrm{CP}}_{[\mathsf{a}]}$ 
 can then be computed explicitly as
\begin{align}
 Z^{\mathrm{CP}}_{[\mathsf{a}]} 
 &=
 e^{-2\pi R_0 E_{\mathrm{GS} } } P_{[a_x, a_y]}
 \nonumber\\
&\quad\times \prod_{s_x\in \mathbb{Z}+a_x} \prod_{s_y\in \mathbb{Z}+a_y} 
\left[ 1 -  e^{-4\pi R_0 \varepsilon(\mathsf{s}) } \right],
\end{align}
where the prefactor $P_{[a_x,a_y]}$ is 
the CP eigenvalue of the ground state (the Fock vacuum).
Note that the partition function does not depend on $a_{\tau}$, which is projected out by CP. 
\footnote{When evaluating the CP twisted partition function (\ref{trace_CP_twisted}), only the simultaneous eigenstates of $H$, $\mathscr{CP}$, and $F$ contribute to the trace.  Since the eigenstates of $\mathscr{CP}$ are charge neutral, it means $e^{2\pi i (a_{\tau}-1/2) F}$ acts as the identity operator inside the trace, and therefore $a_{\tau}$ does not show up in $Z^{\mathrm{CP}}$.
}
In the following, we denote $Z^{\mathrm{CP}}_{[\mathsf{a}]}=Z^{\mathrm{CP}}_{[a_x, a_y]}$ and consider the two cases $a_x=0, 1/2$ separately.

\paragraph{Periodic boundary condition in the $x$ direction, $a_x=0$:}
In this case, we can factor the twisted partition function into the product of 2d massless ($s_z=0$) and massive ($s_x\neq0$) modes
\begin{align}
 &Z^{\mathrm{CP}}_{[a_x=0,a_y]} 
 =
 P_{[a_x=0, a_y]} 
 \\
 &\quad\times
 e^{2\pi R_0 \sum_{s_y}\varepsilon(s_x=0, s_y)  } \prod_{s_y\in \mathbb{Z}+a_y}  \left[ 1 -  e^{-4\pi R_0 \varepsilon(s_x=0, s_y) } \right]
 \nonumber\\
 &\quad\times
 \prod_{s_x\in\mathbb{Z}^+} 
 \left[
 e^{4\pi R_0 \sum_{s_y}\varepsilon(\mathsf{s}) } 
 \prod_{s_y\in \mathbb{Z}+a_y}\left| 1 - e^{-4\pi R_0 \varepsilon(\mathsf{s})} \right|^2
  \right],
  \nonumber
 \end{align}
which can be expressed 
(as the sum $\sum_{s_y}\varepsilon(\mathsf{s})$ is regularized)
 in terms of  
(1+1)d partition functions defined in Sec.\ \ref{Notations}:
 \begin{align}
&Z^{\mathrm{CP}}_{[a_x=0, a_y]} 
=
 P_{[a_x=0, a_y]} e^{ -\pi i (a_y-1/2)} 
 \nonumber\\
 &\qquad\times A^R_{[a_y, 0]} (2i r_{02})
 \prod_{s_x\in\mathbb{Z}^+}
 \Theta_{[a_y, 0]}(2i r_{02}; r_{21}s_x),
 \label{part fun in terms of Kelin and massive}
\end{align}
where $r_{\mu\nu}\equiv R_{\mu}/R_{\nu}$.
When imposing CP symmetry, 
it is reasonable to assume the CP-eigenvalue does not change under $a_y\to a_y+1$,
\cite{Hsieh2014a}
i.e., $P_{[a_x=0,a_y+1]}=P_{[a_x=0,a_y]}$.
Under this assumption, we have
\begin{align}
Z^{\mathrm{CP}}_{[a_x=0,a_y+1]}= -Z^{\mathrm{CP}}_{[a_x=0, a_y]},
\end{align}
where we note
both $A^R_{[a_y, 0]}$ and $\Theta_{[a_y, 0]}$ are invariant under $a_x\to a_x+1$.
The anomalous minus sign under the large gauge transformation,
which comes from the 2d massless modes ($s_x=0$) but not the massive modes ($s_x\neq0$),
signals a $\mathbb{Z}_2$ topological classification:
the CP projected theory can only be realized as the surface theory of a (3+1)d bulk CP symmetric TI, 
which is CPT-conjugate to a (3+1)d time-reversal symmetric TI. \cite{Hsieh2014b}


\paragraph{Antiperiodic boundary condition in the $x$ direction, $a_x=1/2$:}
In this case, the twisted partition function is given by 
\begin{align}
&
 Z^{\mathrm{CP}}_{[a_x=1/2,a_y]} 
 =
  P_{[a_x=1/2, a_y]}
 \prod_{s_x\in\mathbb{Z}^+-1/2}
 \Theta_{[a_y, 0]}(2i r_{02}; r_{21}s_x). 
\end{align}
Observe that there are no 2d massless modes arising in the expression of $Z^{\mathrm{CP}}_{[a_x=1/2,a_y]}$
(while the product of all massive modes changes from $\prod_{s_y\in\mathbb{Z}^+}$ to $\prod_{s_y\in\mathbb{Z}^+-1/2}$).
This partition function is anomaly-free under the large gauge transformation, i.e.,
\begin{align}
\label{LGT_Z^CP_APBC}
Z^{\mathrm{CP}}_{[a_x=1/2,a_y+1]}
= Z^{\mathrm{CP}}_{[a_x=1/2,a_y]}.
\end{align}

 For arbitrary number $N$ of the Dirac fermion flavors, the result is summarized as:
 \begin{align}
\left(Z^{\mathrm{CP}}_{[a_x,a_y+1]}\right)^N= (-1)^{2N(a_x-1/2)}\left(Z^{\mathrm{CP}}_{[a_x=0, a_y]}\right)^N.
\end{align}
The surface theory, as projected (or twisted) by CP, is anomaly-free if and only if $N=0 \mod 2$.
This characterizes the $\mathbb{Z}_2$ classification of the bulk SPT phase.

\section{Surface theory of (3+1)d reflection symmetric crystalline topological superconductors}
\label{Surface theory of (3+1)d reflection symmetric crystalline topological superconductors}

In this section,  
we identify a global gravitational anomaly
of the surface theory of (3+1)d reflection symmetric crystalline TSCs,
which are related to, by CPT-theorem, (3+1)d time-reversal symmetric TSCs. 
While the $\mathbb{Z}_2$-type (gauge) anomaly in the surface of CP TIs agrees with the non-interacting classification of the bulk phase,
the (gravitational) anomaly in the surface of reflection symmetric TSCs, as we will discuss later, sees only the reduction of non-interacting classification, and hence can detect the effect of interactions (in the case that the bulk gap is not destroyed by the interactions).

\subsection{Surface theory}

At the quadratic level, time-reversal symmetric superconductors in  symmetry class DIII
are classified by an integer topological invariant, the 3d winding number $\nu$.
\cite{Schnyder2008}
The topological invariant counts the number of gapless surface Majorana cones.  
For example, the B-phase of $^3\mathrm{He}$
is a TSC (superfluid) with $\nu=1$,
and hosts, when terminated by a surface, 
a surface Majorana cone,
which can be modeled, at low energies, by the Hamiltonian
\begin{align}
 H =
 \frac{1}{(2\pi)^2}
 \int d^2 \mathsf{r}\, \lambda^T (-i \sigma_3 \partial_x- i \sigma_1 \partial_y)\lambda, 
 \label{surf theory}
\end{align}
where
$\sigma_{1,2,3}$ are the Pauli matrices, 
the spatial coordinate $\mathsf{r}=(x,y)\in [0, 2\pi R_1)\times[0,2\pi R_2)$ parameterizes the 2d surface, 
and 
$\lambda(\mathsf{r})$ is a two-component real fermionic field satisfying 
$\lambda^{\dag}(\mathsf{r})=\lambda(\mathsf{r})$.
The surface Hamiltonian is invariant under time-reversal $\mathscr{T}$
defined by 
$
\mathscr{T} \lambda (\mathsf{r}) \mathscr{T}^{-1}  =  i \sigma_2 \lambda(\mathsf{r}),  
$
where 
$\mathscr{T}^2$ is equal to the fermion number parity  
$\mathscr{G}_f=(-1)^F=\frac{1}{(2\pi)^2}\int d^2 \mathsf{r}\,\lambda^T\sigma_2\lambda$.
For TSCs with $\nu=N_f$, the
surface modes can be modeled by $N_f$ copies of the above surface Hamiltonian.

While, at the quadratic level,  , 
one can verify that, for an arbitrary integer $\nu=N_f$,
surface Majorana cones are stable against perturbations
the surface Majorana cones may be destabilized once interactions are included.  
A number of arguments, such as 
``vortex condensation approach'', 
``symmetry-preserving surface topological order'',  
``cobordism approach'' 
and so on,
\cite{Metlitski2014, Fidkowski2013, Kapustin2014c, You2014, KitaevUnpublished}
show that the surface Majorana cones are unstable against interactions 
when $\nu=0$ mod 16, reducing the non-interacting integer classification
to $\mathbb{Z}_{16}$. 

Here, instead of time-reversal symmetry, we consider its CPT-conjugate, reflection or parity symmetry,
which acts on the Majorana field as 
\begin{align}
 \mathscr{P}\lambda(x,y) \mathscr{P}^{-1} = \sigma_3 \lambda(x,2\pi R_2-y),
 \quad
 \mathscr{P}^2=1.   
\label{P_on_psi} 
\end{align}
Upon demanding the invariance under parity (\ref{P_on_psi}),
the Majorana Hamiltonian (\ref{surf theory})
describes the surface of 
symmetry class D + R$_{+}$ crystalline TSCs,
which are, at the quadratic level, classified 
by the integral mirror Chern number. 
\cite{Chiu13, Morimoto13, Shiozaki14} 
Based on CPT-theorem, we expect,  upon the inclusion of interactions, 
the integer classification collapses down to $\mathbb{Z}_{16}$.  

To see the stability of the gapless Majorana mode at the quadratic level, 
note that the mass $\lambda^T \sigma_2 \lambda$ is odd under parity (\ref{P_on_psi})
and prohibited.
It is also interesting to note that 
while the uniform mass is not allowed,  
one could consider  
$
\int d^2 \mathsf{r}\, m(\mathsf{r}) \lambda^T \sigma_2 \lambda
$
with $m(x,2\pi R_2-y) = -m(x,y)$.
This perturbation gaps out the most part of the surface, 
but not completely. 
At the fixed points of $P$ symmetry, 
$y = 0$ and $y=\pi R_2$, 
it leaves gapless modes localized at the domain walls. 
Note that this is similar to the chiral mode localized at 
a mass domain wall on the surface of time-reversal symmetric TIs.
The difference, however, is that 
in the present case, the mass domain wall, as a whole, preserves the reflection symmetry,
while the domain wall on the surface of TIs breaks time-reversal symmetry,
except at the domain wall.
The gapless mode at the domain wall consists of the $N_f$ copies of Majorana fermions 
propagating in either $+x$ or $-x$ directions, depending on the overall sign of the mass domain wall
for each flavor. 
For even  $N_f$, we can always choose (technically) a
set of mass parameters such that the gapless modes at the domain wall are made nonchiral
({\it e.g.}, by choosing different signs of the masses for different flavors of Majorana fermions).
In this case, reflection symmetry acts on the (1+1)d gapless domain-wall fermions as an unitary on-site $Z_2$
symmetry. 
Using the result in Ref.\ \onlinecite{Ryu2012}, it can be shown that such gapless domain-wall states can be gapped without
breaking the symmetry if the number of the non-chiral states is 0 mod 8. This means,
when $N_f = 0$ mod 16, we can gapped out the surface of the crystalline TSCs while preserving the reflection
symmetry at the same time. This gives the $\mathbb{Z}_{16}$ classification, as expected to the same as the case of class DIII TCSs, of the class D + R+ crystalline TSCs, upon the inclusion of interactions.
A similar argument for the $Z_8$ classification of interacting crystalline TIs protected by reflection symmetry
can be found in Ref.\ \onlinecite{Isobe2015}.


\subsection{Projected partition function by reflection/parity symmetry}

We now study the presence/absence of (global) gravitational anomalies of the surface theory (\ref{surf theory}), which signals the existence of the nontrivial bulk SPT phases.
For convenience, we double the degrees of freedom and consider Dirac instead of Majorana fermion fields.  
(This is purely a matter of convenience. The analysis below can be repeated without referring to the Dirac fermion,
and can be done solely in terms of the Majorana fermion.)
The number of Dirac fermion flavors will be denoted by $N$, which corresponds to $N_f = 2N$ in term of the original Majorana fermions.

Our starting point is the partition function
$Z_{[\mathsf{a}]}(g) = \mathrm{Tr}_{a_x, a_y} 
e^{2\pi i (a_{\tau}-1/2) F}
e^{- 2\pi R_0 H'}$ with $H'$ given by Eq.\ (\ref{boosted_H'_T^3}).
Let us now include the effects of parity symmetry by including twisted boundary conditions by parity. 
First, note that 
the modular parameters $\beta$ and $\gamma$ are odd under parity. 
Hence they will be set to zero henceforth,
$\beta=\gamma =0$,
to consider the parity twisted partition function.
While $\mathrm{SL}(3,\mathbb{Z})$ acts on the metric
$g=g(R_i, \alpha, \beta, \gamma)$, 
there is an $\mathrm{SL}(2,\mathbb{Z})$ subgroup generated by $U'_1$ and $U_2$, acting on the 
``reduced'' set of the modular parameters,
$g_{\mathrm{P}}=g_{\mathrm{P}}(R_i,\alpha) \equiv g(R_i, \alpha, \beta= \gamma=0)$.
\footnote{
For Dirac fermions, 
parity also restricts the possible values of the background flux   
to be 
$\mathsf{a}_{\mathrm{P}}\equiv(a_{\tau}, a_x, a_y=0,1/2)$.
However, since our theory here is considered as a double theory for Majorana fermions, $a_{\mu}$ takes only $0$ or $1/2$.}
With the reduced set of modular parameters by parity,  
the total partition function, which is generated by projection by parity and the fermion number parity,  
is given by  
\begin{align}
 \mathcal{Z}^{tot}(g_{\mathrm{P}})
 &=
 \sum_{
 \mathsf{G}\in SG^3 
 }
 \epsilon^{\ }_{\mathsf{G}}
 \, 
\left[ Z_{\mathsf{G}}(g_{\mathrm{P}})\right]^N,
 \nonumber \\
Z_{\mathsf{G}} (g_{\mathrm{P}}) &= \mathrm{Tr}^{\ }_{{G}_x, {G}_y} \left[ {G}_{\tau} (-1)^F e^{-2\pi R_0H'}\right],
 \label{sym_proj_partition function}
\end{align}
where
$SG = \{1,  \mathscr{G}_f, \mathscr{P}, \mathscr{P}\mathscr{G}_f \}$ 
is the symmetry group of the surface fermion theory,
and $\epsilon_{\mathsf{G}}$ are weights assigned to different sectors 
with partition functions twisted by  
$\mathsf{G}=(G_{\tau},{G}_x,{G}_y)$, 
where  for each direction, 
the boundary condition is twisted  
by ${G}_{\mu}=1, \mathscr{G}_f, \mathscr{P}, \mathscr{P}\mathscr{G}_f$.

\begin{figure}[tbp]
\centering
\scalebox{0.35}{\includegraphics{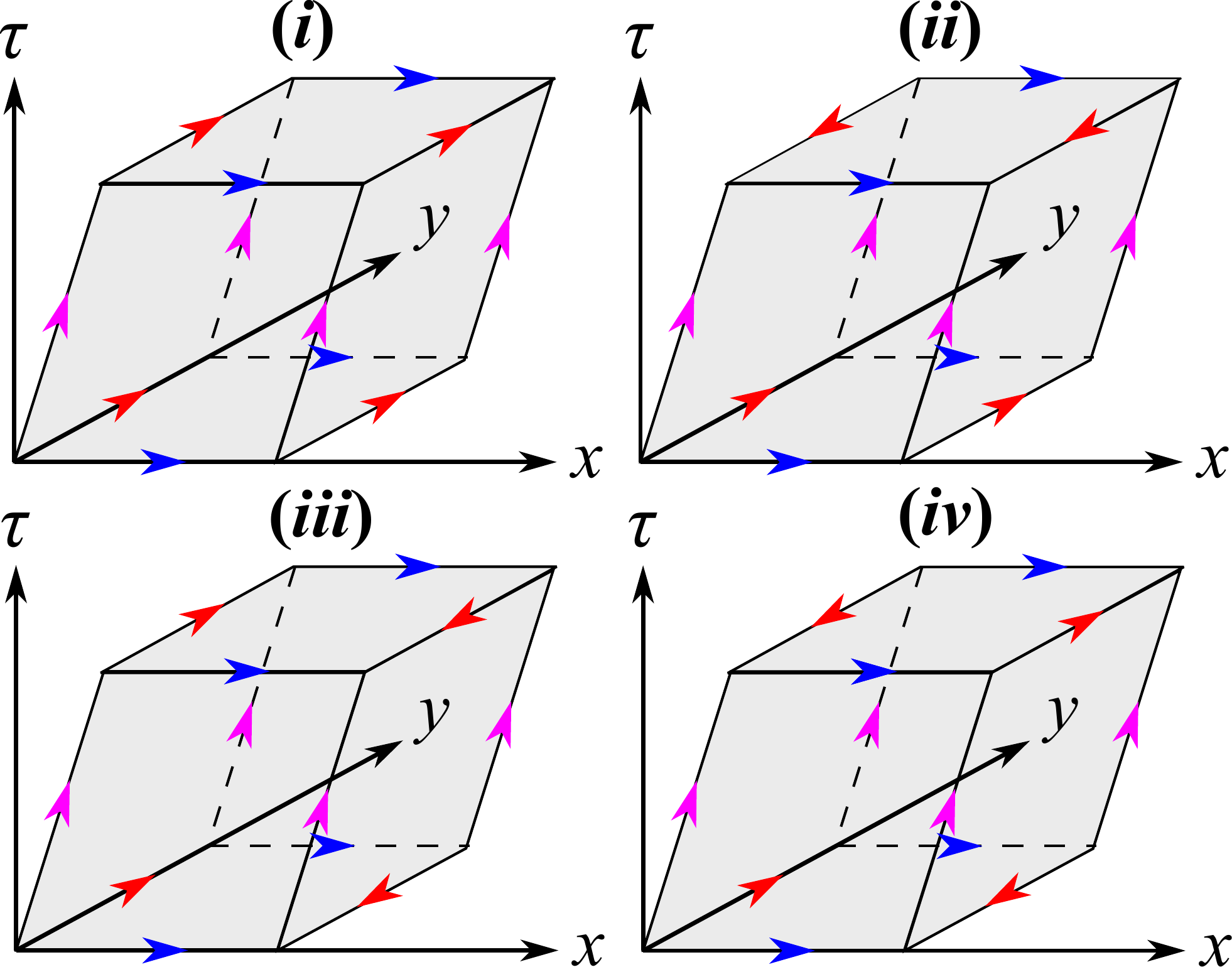}}
\caption{
\label{P_twisted_sectors}
The three-torus and its (unorientable) descendants generated by the orientifold projection. 
While the $y$-boundary condition is twisted by $G_y=1$  or $\mathscr{G}_f$, 
the ${\tau}$- and $x$- boundary conditions are twisted
by 
$({G}_{\tau}, G_{x})=$
$(\mathscr{G}_f^{2a_{\tau}}, \mathscr{G}_f^{2a_{x}})$,
$(\mathscr{P}\mathscr{G}_f^{2a_{\tau}}, \mathscr{G}_f^{2a_{x}})$, 
$(\mathscr{G}_f^{2a_{\tau}}, \mathscr{P}\mathscr{G}_f^{2a_{x}})$,
$(\mathscr{P}\mathscr{G}_f^{2a_{\tau}}, \mathscr{P}\mathscr{G}_f^{2a_{x}})$,
as shown in figures ($i$)--($iv$), respectively. 
(Un)twisted boundary conditions are represented by arrows with the same color.
}

\end{figure}

Not all sectors of the total partition function are mixed by $\mathrm{SL}(2,\mathbb{Z})$. 
We can then divide different sectors into groups,
and study the action of $\mathrm{SL}(2,\mathbb{Z})$ on each group separately. 
In the following, we will focus on the sectors generated by twisting $y$-boundary condition by $1,\mathscr{G}_f$, 
and by twisting ${\tau}$- and $x$- boundary conditions by $1, \mathscr{G}_f, \mathscr{P}, \mathscr{G}_f \mathscr{P}$. 
For a given $y$-boundary condition,
there are $4^2=16$ sectors in total, 
and the corresponding partition function is  
\begin{align}
 \mathcal{Z}^{tot}_{[a_y]}(g_{\mathrm{P}})
 &=
 \sum_{\mathsf{G}\in SG^2}
 \epsilon^{\ }_{\mathsf{G}, a_y}\,
\big[ Z_{(\mathsf{G}, \mathscr{G}_f^{2a_{y}}) } (g_{\mathrm{P}})\big]^N, 
 \label{P_proj_partition function}
\end{align}
where $\mathsf{G}=(G_{\tau}, G_x)$, and $a_y=0,1/2$ represents the $y$-boundary condition. 
We will consider the cases of $a_y=0$ and $a_y=1/2$ separately, 
as they are not mixed by $\mathrm{SL}(2,\mathbb{Z})$. 
The remaining sectors can be generated by twisting $y$-boundary condition by $\mathscr{P}$ and $\mathscr{P}\mathscr{G}_f$. 
Twisting by these group elements gives rise to what can be interpreted as "open sectors" (partition functions on orbifolds) 
as noted by Horava.
\cite{Horava1989}
In this paper, however, we will focus on the 32 ``closed'' sectors generated by twisting with ${G}_y=1, \mathscr{G}_f$.
The resulting closed orientable/unorientable three-manifolds, 
where the (twisted) partition functions are evaluated, are shown in Fig.\ \ref{P_twisted_sectors}.

In the following, we present the analysis of the twisted partition functions for the case of  $a_y=0$. 
The detail of the calculations is left to Appendix\ 
\ref{Parity twisted partition functions of the surface theory of crystalline topological superconductors}.
The analysis for the case of $a_y=1/2$ is similar and in fact simpler. 
In short, for $a_y=1/2$, 
the total partition function (\ref{P_proj_partition function}) 
can be made modular invariant for any  number of Dirac fermion flavors, $N$.
See Appendix\ \ref{Appendix_D}.

On the other hand, 
the total partition function for $a_y=0$ can or cannot be made modular invariant,
depending on $N$. 
For $a_y=0$, 
there are 16 sectors in total, 
generated by twisting by 
$\mathscr{P}$ and $\mathscr{G}_f^{2a_{\mu}}$
in the ${\tau}$ or/and $x$ directions.
We divide these 16 sectors into four sets ($i$--$iv$) 
by
$({G}_{\tau}, G_{x})=$
$(\mathscr{G}_f^{2a_{\tau}}, \mathscr{G}_f^{2a_{x}})$,
$(\mathscr{P}\mathscr{G}_f^{2a_{\tau}}, \mathscr{G}_f^{2a_{x}})$, 
$(\mathscr{G}_f^{2a_{\tau}}, \mathscr{P}\mathscr{G}_f^{2a_{x}})$,
$(\mathscr{P}\mathscr{G}_f^{2a_{\tau}}, \mathscr{P}\mathscr{G}_f^{2a_{x}})$,
respectively.
(There are four sectors in each set.)
The symmetry-twisted partition functions 
$Z_{(G_{\tau}, G_x, \mathscr{G}_f^{2a_y})}$
for each set are then given by
(see Appendix\ \ref{Parity twisted partition functions of the surface theory of crystalline topological superconductors})
\begin{align}
\label{P-twisted_partition functions_sets}
\chi^{i}_{[a_{\tau},a_x]}(g_{\mathrm{P}})
&= 
A^R_{[a_x,a_{\tau}]}(\tau_{2d})A^L_{[a_x,a_{\tau}]}(\tau_{2d})\Theta^{i}_{[a_x,a_{\tau}]},
\\
\chi^{ii}_{[a_{\tau},a_x]}(g_{\mathrm{P}})
&= 
A^R_{[a_x,a_{\tau}]}(\tau_{2d})A^L_{[a_x,a_{\tau}-\frac{1}{2}]}(\tau_{2d})\Theta^{ii}_{[a_x,2a_{\tau}]},
\nonumber\\
\chi^{iii}_{[a_{\tau},a_x]}(g_{\mathrm{P}})
&= 
A^R_{[a_x,a_{\tau}]}(\tau_{2d})A^L_{[a_x-\frac{1}{2},a_{\tau}]}(\tau_{2d})\Theta^{iii}_{[2a_x,a_{\tau}]},
\nonumber\\
\chi^{iv}_{[a_{\tau},a_x]}(g_{\mathrm{P}})
&= 
A^R_{[a_x,a_{\tau}]}(\tau_{2d})A^L_{[a_x-\frac{1}{2},a_{\tau}-\frac{1}{2}]}(\tau_{2d})\Theta^{iv}_{[2a_x,a_{\tau}-a_x]},
\nonumber
\end{align}
where $a_{\tau,x}=0, 1/2$ and we have introduced the functions
$\Theta^{i-iv}_{[a_x,a_{\tau}]}(\tau_{2d}; r_{12})$ by
\begin{align}
\Theta^{i}_{[a_x,a_{\tau}]}
&=
\prod_{s_y\in\mathbb{Z}^+}\left[ \Theta_{[a_x,a_{\tau}]}\left(\tau_{2d} ; r_{12}s_y\right) \right]^2,
\nonumber \\
\Theta^{ii}_{[a_x,a_{\tau}]}
&=
\prod_{s_y\in\mathbb{Z}^+}\Theta_{[a_x,a_{\tau}]}(2\tau_{2d} ; r_{12}s_y),
\nonumber\\
\Theta^{iii}_{[a_x,a_{\tau}]}
&=
\prod_{s_y\in\mathbb{Z}^+}\Theta_{[a_x,a_{\tau}]}(\tau_{2d}/2 ; 2r_{12}s_y),  
\nonumber \\
\Theta^{iv}_{[a_x,a_{\tau}]}
&=
\prod_{s_y\in\mathbb{Z}^+}\Theta_{[a_x,a_{\tau}]}(\tau_{2d}/2+1/2 ; 2r_{12}s_y). 
\label{def_four_Theta}
\end{align}
When evaluating the partition sum (\ref{P_proj_partition function}),
constant prefactors may show up, but are not displayed in the expressions (\ref{P-twisted_partition functions_sets}). 
These prefactors correspond to parity eigenvalues of the ground states in different sectors
[which might depend on the modular parameters and fluxes but are assumed to be $\mathrm{SL}(2,\mathbb{Z})$ invariant],
and can be absorbed to the (redefined) weights 
$\epsilon_{\mathsf{G}, a_y}$ in 
Eq.\ (\ref{P_proj_partition function}).

We now ask, for a specific choice of $N$, by summing these partition functions with some set of weights, 
if we can construct a modular invariant. 
The transformation properties of the twisted partition functions 
$\chi^{i-iv}_{[a_{\tau},a_x]}$
under 
$\mathrm{SL}(2,\mathbb{Z})$ (generated by $U'_1$ and $U_2$)
can be deduced from the properties of $A^{R,L}$ and $\Theta$
shown in Sec. \ref{Notations}; see Appendix\ \ref{Appendix_E}. 
%
It can be shown that if and only if  
$N=4 n$ $(n=1,2,3,\ldots)$,
i.e., 
$N_f=8 n$,  
a modular invariant can be constructed.
In addition,
while $\mathrm{SL}(2,\mathbb{Z})$ invariance can be achieved for $N=4n$,  
there is 
a distinction between $n=2k-1$ and $n=2k$
$(k=1,2,3,\ldots)$,
i.e., 
$N= 8 k-4$ ($N_f=16k-8$)
and
$N=8k$ ($N_f=16k$).
To be explicit, 
the twisted partition functions in set $(i)$ are closed under $\mathrm{SL}(2,\mathbb{Z})$ and 
a modular invariant can be constructed for any $N$. 
For the twisted partition functions in set $(ii-iv)$,
we consider a weighted sum
$\sum_{A=ii,iii,iv} \sum_{i=1}^4 \epsilon^A_i (\chi^{A}_{i})^N$, where
$\chi^{A}_1=\chi^{A}_{[0,0]},\ 
  \chi^{A}_2=\chi^{A}_{[\frac{1}{2},0]},\ 
  \chi^{A}_3=\chi^{A}_{[0,\frac{1}{2}]},\
  \chi^{A}_4=\chi^{A}_{[\frac{1}{2},\frac{1}{2}]}$.
When $N=8k-4$, 
the $\mathrm{SL}(2,\mathbb{Z})$ invariance is achieved when 
\begin{align}
&
(\epsilon^{ii}_1, \epsilon^{ii}_2, \epsilon^{ii}_3, \epsilon^{ii}_4,
\epsilon^{iii}_1, \epsilon^{iii}_2, \epsilon^{iii}_3, \epsilon^{iii}_4,
\epsilon^{iv}_1, \epsilon^{iv}_2, \epsilon^{iv}_3, \epsilon^{iv}_4)
\\
&=(a_1, a_2, a_3, a_3, a_1, a_3, a_2, a_3, -a_1, -a_3, -a_3, -a_2),
\nonumber 
\end{align}
where $a_{i=1,2,3}$ are arbitrary phases (signs). 
Thus, when $N=8k-4$,
the trivial choice, $\epsilon^{A}_i=1$ for all $(A,i)$,
is not allowed. 
When $N=8k$, 
on the other hand, 
the $\mathrm{SL}(2,\mathbb{Z})$ invariance is achieved when 
\begin{align}
&
(\epsilon^{ii}_1, \epsilon^{ii}_2, \epsilon^{ii}_3, \epsilon^{ii}_4,
\epsilon^{iii}_1, \epsilon^{iii}_2, \epsilon^{iii}_3, \epsilon^{iii}_4,
\epsilon^{iv}_1, \epsilon^{iv}_2, \epsilon^{iv}_3, \epsilon^{iv}_4)
\nonumber \\
&=(a_1, a_2, a_3, a_3, a_1, a_3, a_2, a_3, a_1, a_3, a_3, a_2).
\end{align}

\paragraph*{Dimensional reduction to the edge theory of the (2+1)d fermionic SPT phase with $\mathbb{Z}_2\times \mathbb{Z}_2$ symmetry.}

The $\mathrm{SL}(2,\mathbb{Z})$ invariance for $N=4n$ ($N_f=8n$) 
may be understood by taking the limit $R_2 \rightarrow 0$ ($r_{12}\rightarrow\infty$).
In this limit, all massive theta functions become 1 and 
the total partition function constructed here reduces to the form of the (1+1)d partition function projected by 
$\mathbb{Z}_2\times \mathbb{Z}_2$ symmetry (fermion number parity conservation for each chirality),
which is the edge theory of the (2+1)d SPT phase with spin parity conservation.
\cite{Ryu2012}
In the latter case, 
the $\mathrm{SL}(2,\mathbb{Z})$ invariance of the $N_f=8n$ symmetry-projected partition function indicates that 
$8n$ helical Majorana edge modes [in (1+1) dimensions] can be gapped 
without breaking the $\mathbb{Z}_2\times \mathbb{Z}_2$ symmetry. 

\section{Discussion}
\label{Discussion}

We have studied global anomalies on surface theories of (3+1)d topological insulators 
and superconductors.
For CP symmetric TIs, which are related to, by CPT-theorem, time-reversal symmetric TIs, 
there is a global U(1) gauge anomaly if the number of the surface Dirac fermion is odd, characterizing the $\mathbb{Z}_2$ classification of the bulk phase.
For reflection symmetric TSCs, which are related to, by CPT-theorem, class DIII TSCs, a global gravitational anomaly is present in the surface theory
when $N_f \neq 0 \mod 8$. 
The corresponding bulk state is topologically distinct from trivial states of matter
even in the presence of interactions, as far as the bulk gap is not destroyed by the interactions. 
On the other hand, 
the weights $\epsilon_{\mathsf{G}}$, determining the relative weights among partition functions
in different sectors, 
have 16-periodicity as a function of $N_f$.  
Our analysis thus presents an alternative approach to the collapse of the non-interacting classification. 

For the cases where we do not find 
any inconsistency (anomaly), 
i.e., the case of TSCs with $N_f=8$,   
the situation may be more subtle.
First of all, 
the theory may suffer from other forms of inconsistency, which have not been studied here,
and hence  
particular calculations presented in this work does not immediately conclude 
that the corresponding (3+1)d bulk theories are topologically trivial. 
Recall that 
we have not included the partition functions twisted in the $y$ direction by $\mathscr{P}$ and $\mathscr{P}\mathscr{G}_f$ [see comments below Eq.\ (\ref{P_proj_partition function})].

Moreover, we studied the problem of global anomalies
by considering surface theories on $T^3$
(and its descendants generated by the orientifold projection).
Even when the theory is shown to be consistent on $T^3$, it may be anomalous once put on a different three-manifold. 
The situation is better understood for 2d conformal field theories (CFTs), 
where once the consistency of the theories at genus one (torus) is established, 
they can be consistently defined on any (oriented) Riemann surfaces. 
For 3d CFTs, there is no such known fact.  
For this reason, 
our quest for anomalies in the surface theories may not be complete.
Nevertheless, our study on anomalies of 3d massless fermions has shown some interesting and nontrivial results.
\footnote{
The questions addressed here, after the completion of this work, was answered in a recent paper by E. Witten. \cite{Witten2015} 
The $N_f=8$ surface theory of a TSC is actually anomaly-free --  in the traditional sense -- on any 3-manifolds, either orientable or unorientable. However, such surface state indeed suffers from some other inconsistencies. When one considers the problem of anomalies in a more subtle way (than the situation considered in this paper), the anomaly is of order 16 rather than 8.
See the discussion in
Ref.\ \onlinecite{Witten2015}.
}

Finally, it is interesting whether our approach can be related to the gapped surface states of (3+1)d SPT phases that develop symmetry-respecting topological orders.
Such connection is recently investigated in Ref.\ \onlinecite{You2015} in the case of the SU(2) global anomaly. \cite{Witten1982} 
Extending such connection to a generic set of interacting SPT phases is left for future studies.

\acknowledgements 

We thank Liang Fu, Matthew Gilbert, and Edward Witten for useful discussion. 
GYC thanks the financial support from ICMT postdoctoral fellowship and NSF grant 
No. DMR-1408713.
GYC thanks deeply Yuan-Ming Lu for helpful discussion. 
SR is supported by Alfred P. Sloan Research Foundation
and the NSF under Grant No. DMR-1455296.

\appendix 

\section{The Dirac fermion theory on two-torus $T^2$}
\label{The Dirac fermion theory on two torus $T^2$}

In this appendix, we review the modular invariance, the $\mathrm{SL}(2,\mathbb{Z})$ invariance,  
of the Dirac fermion theory on two-torus $T^2$.  

For a flat $T^2$, the zweibein can be factorized as 
\begin{align}
 {e^A}_{\mu} =
 \left(
 \begin{array}{cc}
  R_0 & 0 \\
  0 & R_1 \\
 \end{array}
 \right)
 \left(
 \begin{array}{cc}
  1 & 0 \\
  -\alpha & 1 \\
 \end{array}
 \right)
 =
 \left(
 \begin{array}{cc}
  R_0 & 0 \\
  -\alpha R_1 & R_1 \\
 \end{array}
 \right),
\end{align}
and its inverse is given by
\begin{align}
 {e^{\star}_A}^{\mu}=
  \left(
 \begin{array}{cc}
  \frac{1}{R_0} & \frac{\alpha}{R_0} \\
  0 & \frac{1}{R_1} \\
 \end{array}
 \right),
\end{align}
such that 
${e^A}_{\mu}{e^{\star}_A}^{\nu}={\delta_{\mu}}^{\nu}$ and ${e^A}_{\mu}{e^{\star}_B}^{\mu}={\delta^A}_B$. 
Here $R_0$ and $R_1$ are the radii for the directions $0$ and $1$, and $\alpha$ is related to the angle between the directions $0$ and $1$. 
The Euclidean metric is then given by
\begin{align}
 g_{\mu\nu} 
 &=
 {e^A}_{\mu}{e^B}_{\nu}\delta_{AB}
=
 \left(
 \begin{array}{cc}
  R_0^2+\alpha^2R_1^2 & -\alpha R_1^2 \\
  -\alpha R_1^2 & R_1^2 \\
 \end{array}
 \right),
\end{align}
and the corresponding line element is 
\begin{align}
 ds^2=g_{\mu\nu}d\theta^{\mu}d\theta^{\nu}
 =R_0^2(d\theta^0)^2+R_1^2(d\theta^1-\alpha d\theta^0)^2,
\end{align}
where $0\leq \theta^{\mu}\leq 2\pi$ are angular variables.

The group $\mathrm{SL}(2,\mathbb{Z})$ is generated by two transformations:
\begin{align}
 U_1 =
 \left(
 \begin{array}{cc}
 0 & -1 \\
 1 & 0 \\
 \end{array}
\right),
\quad
 U_2 =
 \left(
 \begin{array}{cc}
 1 & 1 \\
 1 & 0 \\
 \end{array}
\right).
\end{align}
$\mathrm{SL}(2,\mathbb{Z})$ transformations on 
the zweibein and metric are induced by
\begin{align}
 {e^A}_{\mu} 
 &\overset{L}{\longrightarrow} 
 \ {{( e L^T)}^A}_{\mu}
 = {L_{\mu}}^{\rho} {e^A}_{\rho},
 \nonumber\\
 {e^{\star}_A}^{\mu}
 &\overset{L}{\longrightarrow} 
 {{(e^{\star}L^{-1})}_A}^{\mu}
 =  {e^{\star}_A}^{\rho}{{(L^{-1})}_{\rho}}^{\mu},
 \nonumber\\
 g_{\mu\nu}  
 &\overset{L}{\longrightarrow} 
 \ {( L g L^T)}_{\mu\nu}
 = {L_{\mu}}^{\rho} {L_{\nu}}^{\sigma}  g_{\rho\sigma},
\end{align}
for any $\mathrm{SL}(2,\mathbb{Z})$ element $ L = U^{n_1}_1 U^{n_2}_2 U^{n_3}_1\cdots$.
In particular,
\begin{align}
 g_{\mu\nu}  
 &\overset{U_1}{\longrightarrow} 
 {( U_1 g U^T_1)}_{\mu\nu} 
=
 \left(
 \begin{array}{cc}
  R_1^2 & \alpha R_1^2 \\
  \alpha R_1^2 & R_0^2+\alpha^2R_1^2 \\
 \end{array}
 \right),
\end{align}
which corresponds to the changes
\begin{align}
R_0 \rightarrow R_0/|{\tau_{2d}}|, \quad
&R_1 \rightarrow R_1|{\tau_{2d}}|, \quad
\alpha \rightarrow -\alpha/|{\tau_{2d}}|^2, 
\end{align}
or, in terms of the modular parameter (the Teichm\"uller parameter) ${\tau_{2d}}\equiv\alpha+i\frac{R_0}{R_1}$, 
\begin{align}
{\tau_{2d}} \rightarrow -1/\tau_{2d}.
\end{align}
On the other hand,  
\begin{align}
 g_{\mu\nu}  
 &\overset{U_2}{\longrightarrow} 
 \ {( U_2 g U^T_2)}_{\mu\nu} 
 \nonumber\\
 &\qquad=
 \left(
  \begin{array}{cc}
  R_0^2+(\alpha-1)^2R_1^2 & -(\alpha-1) R_1^2 \\
  -(\alpha-1) R_1^2 & R_1^2 \\
 \end{array}
 \right),
\end{align}
which corresponds to the change
\begin{align}
\alpha \rightarrow \alpha-1 \quad
\text{(while $R_0$ and $R_1$ are unchanged)}.
\end{align}
The two transformations $U_1$ and $U_2$ are exactly $S$ and $T^{-1}$ transformations that generate $\mathrm{SL}(2,\mathbb{Z})$ 
(usually used in the 2d conformal field theory literatures), respectively.

The Euclidean action for the Dirac fermion on this two torus is given by 
\begin{align}
 S_E
 &= \frac{1}{2\pi}\int d^2 \theta\ \left(\det{e}\right) \bar{\psi} \left(\Gamma^A {e^{\star}_A}^{\mu} \partial_{\theta^{\mu}} \right) \psi,
 \end{align}
where $\det{e}=\sqrt{g}=R_0R_1$, $\partial_{\theta^{\mu}}\equiv \frac{\partial}{\partial\theta^{\mu}}$, 
and  the gamma matrices $\Gamma^A$ satisfy $\{\Gamma^A, \Gamma^B\}= 2\delta^{AB}$.
In terms of the space-time coordinates ${\tau}=R_0\theta^0$ and $x=R_1\theta^1$: 
\begin{align}
 2\pi S_E 
 &= 
 \int_0^{2\pi} d\theta^0 \int_0^{2\pi} d\theta^1
 \nonumber\\
 &\quad
 \times 
 \bar{\psi} 
 \left(R_1\Gamma^0\partial_{\theta^0}+\alpha R_1\Gamma^0\partial_{\theta^1}+R_0\Gamma^1\partial_{\theta^1}\right) \psi 
 \nonumber\\
 &= \int_0^{2\pi R_0} d{\tau} \int_0^{2\pi R_1} dx 
 \nonumber\\
 &\quad
\times
\bar{\psi} 
 \left(\Gamma^0\partial_{{\tau}}+ \alpha\frac{R_1}{R_0}\Gamma^0\partial_x+ \Gamma^1\partial_x\right) \psi.
 \end{align}

The partition function can be evaluated by the path integral on the (general) two torus 
$Z(g)=\int \mathcal{D}[\psi^{\dag},\psi]  e^{-S_E}$,
 or by the operator formalism 
 \begin{align}
 Z(g)=\mathrm{Tr} \left[e^{-2\pi R_0H'}\right],
 \end{align}
 where $H'$ is the "boosted" Hamiltonian (in the presence of non-vanishing $\alpha$) corresponding to $S_E$:\begin{align} 
 H'=  H - i\alpha\frac{R_1}{R_0} P_x,
 \end{align}
with
\begin{align}
 H 
 &=
 \frac{1}{2\pi} 
 \int^{2\pi R_1}_0 dx\, 
 \bar{\psi} \Gamma^1\partial_x\psi, 
 \nonumber\\
 P_x 
 &=
 \frac{1}{2\pi} 
 \int^{2\pi R_1}_0 dx\, 
 \psi^{\dagger}(- i\partial_x\psi)
\end{align}
being the Hamiltonian and momentum on a ''flat two torus" ($\alpha=0$).

The modular invariance for the partition function of nonchiral fermions 
is achieved by summing twisted partition functions over the spin structures.
We thus consider the partition function
 \begin{align}
 \mathcal{Z}^{tot}(g)
 &=
 \sum_{({G}_{{\tau}}, {G}_x)\in SG^2}
 \mathrm{Tr}_{{G}_x}
 \left[{G}_{{\tau}} (-1)^{F} e^{-2\pi R_0H'}\right],
\end{align}
where 
 $SG
 =
\{1, (-1)^{F}\}$
is the symmetry group of 
the free fermion theory. 
Then,
the total partition function 
satisfies 
$
\mathcal{Z}^{tot}(LgL^T)=
\mathcal{Z}^{tot}(g)
$
for 
$
 L\in \mathrm{SL}(2,\mathbb{Z}).
$
\cite{Mirror}

\section{Regularization of the ground state-energy}
\label{Regularization of the ground-state energy}

In this appendix, we regularize the ground-state energy, which is given by the divergent sum
\begin{align}
E_{\mathrm{GS}[\mathsf{a}]}(g)
=
-\sum_{\mathsf{s}\in\mathbb{Z}^2+(a_x, a_y)}|\mathsf{s}|,
\end{align}
where $|\mathsf{s}|\equiv\sqrt{g_2^{ij}s_is_j}$.

Following Appendix C in Ref.\ \onlinecite{Dolan2013}, for arbitrary positive integer $d$, we have

\begin{align}
\label{regularized_sum}
\sum_{\mathsf{s}\in\mathbb{Z}^d+\mathsf{\alpha}}|\mathsf{s}|e^{i\mathsf{s}\cdot\mathsf{x}}
= \frac{c_{d+1}}{(2\pi)^d}\sqrt{g_d}\int d^dy 
\frac{1}{|\mathsf{y}|^{d+1}}\sum_{\mathsf{s}}e^{i\mathsf{s}\cdot(\mathsf{x}-\mathsf{y})}, 
\end{align}
where $\mathsf{s}, \mathsf{\alpha}\in\mathbb{R}^d$, $|\mathsf{s}|\equiv\sqrt{g_d^{ij}s_is_j}$,
$g_d\equiv\det(g_{dij})$,
and $c_{d+1}\equiv\frac{\pi^{\frac{d}{2}}2^{d+1}\Gamma\left(\frac{d+1}{2}\right)}{\Gamma\left(\frac{1}{2}\right)}$.
Now we use the equality
\begin{align}
\sum_{\mathsf{m}\in\mathbb{Z}^d}e^{i\mathsf{m}\cdot(\mathsf{x}-\mathsf{y})} 
= (2\pi)^d \sum_{\mathsf{n}\in\mathbb{Z}^d} \delta^d(\mathsf{x}-\mathsf{y}+2\pi\mathsf{n}).
\end{align}
Substituting the above equality, with removing the $\mathsf{n}= 0$ term in the sum, into (\ref{regularized_sum}), we obtain the regularized sum
\begin{align}
&\sum_{\mathsf{s}\in\mathbb{Z}^d+\mathsf{\alpha}}|\mathsf{s}|e^{i\mathsf{s}\cdot\mathsf{x}}
\nonumber\\
&= c_{d+1}\sqrt{g_d}\int d^dy 
\frac{1}{|\mathsf{y}|^{d+1}}\sum_{\mathsf{n}\neq0\in\mathbb{Z}^d} \delta^d(\mathsf{x}-\mathsf{y}+2\pi\mathsf{n}) 
e^{i\mathsf{\alpha}\cdot(\mathsf{x}-\mathsf{y})}
\nonumber\\
&= c_{d+1}\sqrt{g_d} \sum_{\mathsf{n}\neq0\in\mathbb{Z}^d} \frac{e^{-2\pi i\mathsf{\alpha}\cdot\mathsf{n}}}{\left| \mathsf{x}+2\pi\mathsf{n} \right|^{d+1}}
\end{align}
Then our regularized ground-state energy is given by 
\begin{align}
E_{\mathrm{GS}[\mathsf{\alpha}]}(g)
&= -\sum_{\mathsf{s}\in\mathbb{Z}^2+(a_x, a_y)}|\mathsf{s}| 
\left.e^{i\mathsf{s}\cdot\mathsf{x}}\right|_{\mathsf{x}=0}
\nonumber\\
&= -c_3\sqrt{g_2}\sum_{\mathsf{n}\neq0\in\mathbb{Z}^2} 
\frac{e^{-2\pi ia^i n_i}}{\left| 2\pi\mathsf{n} \right|^{3}}
\nonumber\\
&= \frac{1}{4\pi^2}\sqrt{\det(g_{2 ij})}\sum_{\mathsf{n}\neq0\in\mathbb{Z}^2} 
\frac{\cos\left(2\pi ia^i n_i\right)}{\left(g_2^{ij}n_in_j \right)^{\frac{3}{2}}}.
\end{align}

\section{Derivation of the claim (\ref{SL(3,Z)_invariance})}
\label{Derivation of the claim (ref{SL(3,Z)_invariance})}

In this Appendix, we confirm the claim (\ref{SL(3,Z)_invariance})
by explicitly checking how $Z_{[\mathsf{a}]}(g)$ transforms under the two generators $U_1=U'_1M$ and $U_2$ of $\mathrm{SL}(3,\mathbb{Z})$, defined in Eqs.\ (\ref{U1 and U2}) and (\ref{U1' and M}).

The behavior of $Z_{[\mathsf{a}]}(g)$ under $U_2$ and $U'_1$ can be directly deduced 
by the properties of the massive theta function listed in
(\ref{massive_theta_func_properties}). 

\paragraph{Transformation under $U_2$:}

Under $U_2$,
the metric is transformed as in (\ref{mod_under_U_2}), 
while the fluxes are transformed as 
$(a_{\tau}, a_x, a_y)\rightarrow(a_{\tau}+a_x, a_x, a_y).$
From (\ref{full part fn_2}), we have
\begin{align}
&Z_{[U_2 \mathsf{a}]}(U_2g{U_2}^T)
\nonumber\\
&= 
 \prod_{s_{y}\in\mathbb{Z}+a_y} 
  \Theta_{[a_{x}+\beta s_y, a_{\tau}+a_x+(\gamma+\beta) s_y]} 
 \left(\tau_{2d}-1; r_{12} s_y \right)
\nonumber\\
&= 
 \prod_{s_{y}\in\mathbb{Z}+a_y} 
  \Theta_{[a_x+\beta s_y,a_{\tau}+\gamma s_y]} 
\left(\tau_{2d}; r_{12} s_y 
\right)
\nonumber\\
&= 
Z_{[\mathsf{a}]}(g)
\end{align}

\paragraph{Transformation under $U'_1$:}

Under $U_1'$,
the metric is transformed as in (\ref{mod_under_U_1'}), 
while the fluxes are transformed as 
$(a_{\tau}, a_x, a_y)\rightarrow(-a_x, a_{\tau}, a_y).$
From (\ref{full part fn_2}), we have
\begin{align}
&Z_{[U'_1 \mathsf{a}]}(U'_1g{U'_1}^T)
\nonumber\\
&= 
 \prod_{s_{y}\in\mathbb{Z}+a_y} 
  \Theta_{[a_{\tau}+\gamma s_y, -a_x - \beta s_y]} 
\left(-1/\tau_{2d}; r_{12} s_y |\tau_{2d}|
\right)
\nonumber\\
&= 
 \prod_{s_{y}\in\mathbb{Z}+a_y} 
  \Theta_{[-a_x - \beta s_y,-a_{\tau}-\gamma s_y]} 
\left(\tau_{2d}; r_{12} s_y 
\right)
\nonumber\\
&= 
 \prod_{s_{y}\in\mathbb{Z}+a_y} 
  \Theta_{[a_x + \beta s_y,a_{\tau}+\gamma s_y]} 
\left(\tau_{2d}; r_{12} s_y 
\right)
\nonumber\\
&= 
Z_{[\mathsf{a}]}(g)
\end{align}

\paragraph{Transformation under $M$:}

Transformation for the parameters $\{R_i, \alpha, \beta, \gamma\}$ under $M$ 
is not as obvious as the cases of $U_2$ and $U_1'$.
We observe that,
since the transformation $M$ only involves the change in the $x$-$y$ plane,  
under $M$ the $x$- and $y$- components of the dreibein ${e^A}_{\mu}$ 
and the metric $g_{\mu\nu}$ (and their inverses) transform as:
\begin{align}
{e^A}_{i} \rightarrow {M_i}^k{e^A}_{k},
&\quad
{e^{\star}_A}^{i} \rightarrow {e^{\star}_A}^{k}{(M^{-1})_k}^i,
\\
g_{ij} \rightarrow {M_i}^k{M_j}^lg_{kl}, 
&\quad
(g_2)^{ij} \rightarrow {(M^{-1})_k}^i{(M^{-1})_l}^j(g_2)^{kl},
\nonumber
\end{align}
where $i, j, k, l =1, 2$, and $(g_2)^{ij}$ is defined in Eq.\ (\ref{g2_inverse}).
To see the behavior of Eq.\ (\ref{full part fn}) under $M$, 
we first note the regularized ground state energy  
(\ref{GS regularized}) 
satisfies $E_{\mathrm{GS}[M\mathsf{a}]}(MgM^T)=E_{\mathrm{GS}[\mathsf{a}]}(g)$.
On the other hand,
the second line in Eq.\ (\ref{full part fn}) 
can be expressed as ($i, j=1,2$)
\begin{align}
&\quad
e^{
-2\pi R_0 \varepsilon(\mathsf{s})+2\pi i\alpha s_x+2\pi i(\alpha\beta+\gamma)s_y+2\pi ia_{\tau}
}
\nonumber\\
&=
e^{-2\pi R_0\sqrt{g_2^{ij}s_is_j}+2\pi iR_0{e^{\star}_0}^{i}s_i+2\pi ia_{\tau}},
\end{align}
where ${e^{\star}_0}^{i}=(\alpha/R_0, (\alpha\beta+\gamma)/R_0)^T$. 
From this expression, we can see that the mode-product term in 
Eq.\ (\ref{full part fn}) is also invariant under 
$\{g, \mathsf{a}\}\rightarrow \{MgM^T, M\mathsf{a}\}$. 
Therefore, we have shown 
\begin{align}
Z_{[M\mathsf{a}]}(MgM^T)=Z_{[\mathsf{a}]}(g).
\end{align}

From the above discussion, 
we thus confirm our claim (\ref{SL(3,Z)_invariance}).


\section{Parity twisted partition functions of the surface theory of 
crystalline topological superconductors}
\label{Parity twisted partition functions of the surface theory of 
crystalline topological superconductors}

In this Appendix, 
we explicitly calculate the partition functions twisted by parity, 
which is defined by
 \begin{align}
\mathscr{P} \psi(x,y) \mathscr{P}^{-1} = \sigma_3 \psi(x,-y),\quad
\mathscr{P}^2=1,
\end{align}
where $\psi$ is the two-component Dirac fermion.
(Remember that we have doubled the degree of freedom of the original theory of Majorana fermions.)
Here we define $y\rightarrow -y$ instead $y\rightarrow 2\pi R_2-y$ (defined in the main text) by parity is just for convenience (the result does not depend on the choice). 
As mentioned in the text, the parity invariance $\mathscr{P}H'\mathscr{P}^{-1}=H'$ 
forces strictly $\beta=\gamma =0$.
Then, P acts on the Fourier components of the original fermion operators as 
\begin{align}
 \mathscr{P}
\tilde{\psi}(\mathsf{s})
\mathscr{P}^{-1}
=
\sigma_3
\tilde{\psi}({\bar{\mathsf{s}}}), 
\end{align}
where $\bar{\mathsf{s}}= (s_x, -s_y)$.
On the other hand, the P action on the eigen basis $\chi^{\ }_{\pm}$, defined in Eq. (\ref{fermion_eigenbasis}), is deduced as 
\begin{align}
&\quad
\mathscr{P}
\chi(\mathsf{s})
\mathscr{P}^{-1}
\nonumber \\
& 
=
\left[
\begin{array}{cc}
\langle {u}_{+}(\mathsf{s})|\sigma_3  |{u}_+(\bar{\mathsf{s}})\rangle & 
\langle {u}_{+}(\mathsf{s})|\sigma_3 |{u}_-(\bar{\mathsf{s}})\rangle  \\
\langle {u}_{-}(\mathsf{s})|\sigma_3 |{u}_+(\bar{\mathsf{s}})\rangle  
&
\langle {u}_{-}(\mathsf{s})|\sigma_3 |{u}_-(\bar{\mathsf{s}})\rangle   \\
\end{array}
\right]
\chi(\bar{\mathsf{s}}). 
\end{align}
where 
$|{u}_{\pm}(\mathsf{s})\rangle$ are eigenvectors of 
\begin{align}
\mathcal{H}'(\mathsf{s}) = \frac{s_x}{R_1}\sigma_3 + \frac{s_y}{R_2}\sigma_1 + \alpha \frac{s_x}{R_0} 
\end{align}
with eigenvalues $\pm \varepsilon(\mathsf{s})+\alpha s_x/R_0$,
where $\varepsilon(\mathsf{s})=\sqrt{\left(s_x/R_1\right)^2+\left(s_y/R_2\right)^2 }$.
Because of P symmetry,
$\sigma_3\mathcal{H}'(\mathsf{s})U\sigma_3^{-1}=\mathcal{H}'(\bar{\mathsf{s}})$, 
$\sigma_3|{u}_{\pm}(\bar{\mathsf{s}})\rangle$ are also 
eigenvectors of $\mathcal{H}'(\mathsf{s})$ with eigenvalues $\pm \varepsilon(\mathsf{s})+ \alpha s_x/R_0$, 
and therefore the off-diagonal matrix elements are zero, 
$\langle {u}_{+}(\mathsf{s})|\sigma_3|{u}_-(\bar{\mathsf{s}})\rangle=\langle {u}_{-}(\mathsf{s})|\sigma_3|{u}_+(\bar{\mathsf{s}})\rangle=0$.

The diagonal elements, and hence, the transformation properties of $\chi_{\pm}(\mathsf{s})$ under parity, 
depend on a choice of eigen functions $\vec{u}_{\pm}(\mathsf{s})$.  
For $s_y\neq 0$,
the following choice for the eigenvectors: 
\begin{align}
  |u_{\pm}(\mathsf{s}) \rangle
  &=
  \frac{1}{\sqrt{2\varepsilon(\mathsf{s})\left[\varepsilon(\mathsf{s})\pm s_x/R_1\right]}}
  \left[
    \begin{array}{c}
      s_x/R_1\pm \varepsilon(\mathsf{s}) \\
      s_y/R_2
    \end{array}
  \right]
  \label{eigenvectors_u_pm_1_App}
\end{align}
leads to  
\begin{align}
 \langle u_{+}(\mathsf{s}) |\sigma_3 |u_+ (\bar{\mathsf{s}})\rangle 
 &=
 \langle u_{-}(\mathsf{s}) |\sigma_3 |u_- (\bar{\mathsf{s}})\rangle 
 =1.
\end{align}
Alternatively, a different gauge choice 
\begin{align}
  |u_{\pm}(\mathsf{s}) \rangle
  &=
  \frac{1}{\sqrt{2\varepsilon(\mathsf{s})\left[\varepsilon(\mathsf{s})\pm s_x/R_1\right]}}
  \left[
    \begin{array}{c}
      s_y/R_2\\
      -s_x/R_1\pm \varepsilon(\mathsf{s})
    \end{array}
  \right]
 \label{eigenvectors_u_pm_2_App}
\end{align}
leads to 
\begin{align}
 \langle u_{+}(\mathsf{s}) |\sigma_3 |u_+ (\bar{\mathsf{s}})\rangle 
 &=
 \langle u_{-}(\mathsf{s}) |\sigma_3 |u_- (\bar{\mathsf{s}})\rangle 
 =-1.
\end{align}
In either choice, the result can be summarized as
\begin{align}
 \mathscr{P} 
 \left[
 \begin{array}{c}
 \chi_{+}(\mathsf{s}) \\
 \chi_{-}(\mathsf{s}) 
 \end{array}
 \right] 
 \mathscr{P}^{-1} 
 = 
  \left[
 \begin{array}{c}
 \eta_+\chi_{+}(\bar{\mathsf{s}}) \\
 \eta_-\chi_{-}(\bar{\mathsf{s}}) 
 \end{array}
 \right],
 \quad
 s_y\neq 0,
 \end{align}
 where $\eta_{\pm}$ is an $s$-independent sign factor.  
Note that the condition $\mathscr{P}^2=1$ forces $\eta_{\pm}^2=1$. 
While $\eta_{\pm}$ depends on the choice of eigenfunctions, 
the final results (such as the evaluation of the partition functions) do not depend on 
such ambiguity.


On the parity-invariant line $s_y=0$, which exists if $a_y\in\mathbb{Z}$, the Hamiltonian $\mathcal{H}'(s_x, s_y=0)=\frac{s_x}{R_1}\sigma_3 + \alpha\frac{s_x}{R_0}$ has a "chiral decomposition":
\begin{align}
  |u_{R}(s_x) \rangle
  = \left[
    \begin{array}{c}
      e^{i\alpha_R} \\
      0
    \end{array}
  \right],
  \quad
  |u_{L}(s_x) \rangle
  = \left[
    \begin{array}{c}
      0 \\
      e^{i\alpha_L} 
    \end{array}
  \right],
  \quad
  \alpha_{R,L}\in\mathbb{R},
  \label{eigenvectors_u_RL_App}
\end{align}
which corresponds to the "chiral eigen basis" $\chi_{R,L}$.
Since $\langle u_{R}(s_x)|\sigma_3|u_{R}(s_x)\rangle=-\langle u_{L}(s_x)|\sigma_3|u_{L}(s_x)\rangle=1$ and $\langle u_{R}(s_x)|\sigma_3|u_{L}(s_x)\rangle=\langle u_{L}(s_x)|\sigma_3|u_{R}(s_x)\rangle=0$ (independent of the choice of the phases $\alpha_{R/L}$), we have
\begin{align}
 \mathscr{P} 
 \left[
 \begin{array}{c}
 \chi_{R}(s_x) \\
 \chi_{L}(s_x) 
 \end{array}
 \right] 
 \mathscr{P}^{-1} 
 = \sigma_3
  \left[
 \begin{array}{c}
 \chi_{R}(s_x) \\
 \chi_{L}(s_x) 
 \end{array}
 \right],
 \quad
 s_y= 0,
 \end{align}
which does not depend on the normalizations of $|u_{R/L}(s_x)\rangle$.
We observe, 
on the P-invariant line $s_y=0$, 
parity acts like the "spin parity" $(-1)^{F_L}$, 
where $F_L$ can be thought as the total number of $\chi_{L}(s_x)$ (at $s_y=0$). 
Thus, we expect that the modular properties of this surface theory,
as determined solely by the 2d massless modes $(s_x, s_y=0)$,
will be similar to the modular properties of the edge theory of 
(2+1)d topological superconductors protected by $\mathbb{Z}_2\times\mathbb{Z}_2$ symmetries.
\cite{Ryu2012}

\paragraph{P-twisted partition functions in the ${\tau}$-direction}

First we evaluate the partition function twisted by P in the ${\tau}$-direction,
which can be written as
\begin{align}
&\quad 
Z_{
\mathscr{P}\mathscr{G}^{2a_{\tau} }_f, 
\mathscr{G}^{2a_x}_f, a_y 
}
\nonumber \\
&=
\mathrm{Tr}_{\mathscr{G}_f^{2a_x},a_y}
[\mathscr{P} e^{2\pi i (a_{\tau}-1/2)F} 
e^{-2\pi R_0H'}] 
 \nonumber \\
 &=
 e^{ -2\pi R_0 E_{\mathrm{GS}}}
 \prod_{s_x} W^{\mathrm{P}}_{[\mathsf{a}]}(s_x),
 \label{P_twisted_partition functions_tau_direction_appendix}
\end{align}
where $a_y=0,1/2$,
$E_\mathrm{GS} = - \sum_\mathsf{s} \varepsilon(\mathsf{s})$,
and
$W^{\mathrm{P}}_{[\mathsf{a}]}(s_x)$ can be written in a pairwise fashion (with respect to P symmetry):
\begin{align}
&W^{\mathrm{P}}_{[\mathsf{a}]}(s_x)
=
X^{\mathrm{P}}_{[\mathsf{a}]}(s_x)
\times 
Y^{\mathrm{P}+}_{[\mathsf{a}]}(s_x)
\times
Y^{\mathrm{P}-}_{[\mathsf{a}]}(s_x),
\end{align}
where
\begin{align}
&
X^{\mathrm{P}}_{[\mathsf{a}]}(s_x)
=
\mathrm{Tr}_{a_xa_y}\,
 \mathscr{P}
 \exp\Big\{
 \nonumber \\
 &\qquad 
+ 2\pi i \Big(a_{\tau}-\frac{1}{2}\Big)
 \left[\chi^{\dag}_{R}(s_x) \chi^{\ }_{R}(s_x) + \chi^{\dag}_{L}(s_x) \chi^{\ }_{L}(s_x)\right]
 \nonumber\\
 &\qquad
 -2\pi s_x \frac{R_0}{R_1}
 \left[\chi^{\dag}_{R}(s_x) \chi^{\ }_{R}(s_x) - \chi^{\dag}_{L}(s_x) \chi^{\ }_{L}(s_x)\right]
 \nonumber \\
 &\qquad 
 +2\pi i\alpha s_x\left[\chi^{\dag}_{R}(s_x) \chi^{\ }_{R}(s_x) + \chi^{\dag}_{L}(s_x) \chi^{\ }_{L}(s_x)\right]
 \Big\}
 \end{align}
 and 
 \begin{align}
&
Y^{\mathrm{P}\pm }_{[\mathsf{a}]}(s_x)
=
 \prod_{s_y>0}
 \mathrm{Tr}_{a_xa_y}\,
 \mathscr{P}
 \exp\Big\{
 \nonumber \\
 &\qquad 
 \pm 2\pi i\Big(a_{\tau}-\frac{1}{2}\Big)
 \left[\chi^{\dag}_{\pm}(\mathsf{s}) \chi^{\ }_{\pm}(\mathsf{s}) 
 + \chi^{\dag}_{\pm}(\bar{\mathsf{s}}) \chi^{\ }_{\pm}(\bar{\mathsf{s}})\right]
 \nonumber\\
 &
 \quad\quad
 -2\pi \varepsilon(\mathsf{s}) \frac{R_0}{R_1}
 \left[\chi^{\dag}_{\pm}(\mathsf{s}) \chi^{\ }_{\pm}(\mathsf{s}) 
 + \chi^{\dag}_{\pm}(\bar{\mathsf{s}}) \chi^{\ }_{\pm}(\bar{\mathsf{s}})\right]
 \nonumber \\
 &\qquad 
 \pm 2\pi i\alpha s_x
 \left[
 \chi^{\dag}_{\pm}(\mathsf{s}) \chi^{\ }_{\pm}(\mathsf{s}) 
 + \chi^{\dag}_{\pm}(\bar{\mathsf{s}}) \chi^{\ }_{\pm}(\bar{\mathsf{s}})\right]
 \Big\}.
\end{align}
Note that the 2d massless modes ($s_y=0$) $X^{\mathrm{P}}_{[\mathsf{a}]}(s_x)$ would be present if $a_y=0$.
With such pairwise decomposition, 
the 2d massive part for fixed $s_y\neq 0$ in Eq.\ (\ref{P_twisted_partition functions_tau_direction_appendix}) 
is evaluated as 
\begin{align}
 &e^{ 2\pi R_0 \sum_{s_x\in \mathbb{Z}+a_x}\varepsilon(\mathsf{s})}
\prod_{s_x\in \mathbb{Z}+a_x} 
 \left|
 1-e^{-2\pi R_0\varepsilon(\mathsf{s})+2\pi i\alpha s_x+2\pi i a_{\tau}}
 \right|^2
 \nonumber\\
 &=
 \Theta_{[a_x, 2a_{\tau}] } \left({\tau_{2d}}; r_{12} s_y\right),
\end{align}
while the  2d massless part is evaluated as 
\begin{align}
 A^R_{ [a_x,a_{\tau}]}({\tau_{2d}})A^L_{ [a_x,a_{\tau}-\frac{1}{2}]}({\tau_{2d}}_{}).
\end{align}
In summary, 
\begin{align}
&
Z_{
\mathscr{P}\mathscr{G}^{2a_{\tau} }_f, \mathscr{G}^{2a_x}_f, a_y
}
\nonumber \\
&=
\left\{
\begin{array}{l}
\text{const.} \times
A^R_{[a_x,a_{\tau}]}({\tau_{2d}})A^L_{ [a_x,a_{\tau}-\frac{1}{2}]}({\tau_{2d}}_{}) 
 \nonumber\\
 \displaystyle 
\qquad \times  \prod_{s_y\in\mathbb{Z}^+}  \Theta_{[a_x, 2a_{\tau}] } \left(2{\tau_{2d}}; r_{12} s_y\right)
\\
\qquad \mbox{for $a_{y}=0$ (PBC in the $y$-direction)}, 
\\
\\
\\
\displaystyle 
\text{const.} \times
 \prod_{s_y\in\mathbb{Z}^+-\frac{1}{2}}  
 \Theta_{[a_x, 2a_{\tau}] } \left(2{\tau_{2d}}; r_{12} s_y\right)
 \\
\qquad \mbox{for $a_{y}=1/2$ (APBC in the $y$-direction)}. 
 \end{array}
\right.
\end{align}
Here the constant prefactors are related to the P eigenvalues of the ground states.

\paragraph{P-twisted partition functions in the $x$-direction}

Now let us consider the partition function twisted by P in the $x$-direction.
We start with the twisted boundary conditions in the $x$- and $y$-directions:
\begin{align}
 \psi(x+2\pi R_1,y) 
 &= 
\big(\mathscr{P}\mathscr{G}_f^{2a_x}\big)
 \psi(x, y) 
 \big(\mathscr{P}\mathscr{G}_f^{2a_x}\big)^{-1}
 \nonumber\\
 &=
 e^{ 2\pi i a_x} \sigma_3\psi(x,-y),
\nonumber \\
 \psi(x,y+2\pi R_2) 
 &= 
\big(\mathscr{G}_f^{2a_x}\big)\psi(x, y) (\mathscr{G}_f^{2a_x}\big)^{-1}
 \nonumber\\
 &=
 e^{ 2\pi i a_y}\psi(x,y).
\end{align}
With the above twisted boundary condition, 
the Fourier expansion of the fermion fields can be expressed as 
\begin{align}
 \psi(\mathsf{r}) 
 &= 
 \sum_{s_x\in \mathbb{Z}/2+a_x} \sum_{s_y \in \mathbb{Z}+ a_y}
 \nonumber\\
 &\qquad
 e^{  i x \frac{s_x}{R_1}+ i y \frac{s_y}{R_2}} 
 \left[
 \vec{u}_+(\mathsf{s})\chi_+(\mathsf{s})+\vec{u}_-(\mathsf{s})\chi_-(\mathsf{s})
 \right]
 \nonumber\\
 &\quad+
 \mbox{\{2d massless modes\}},
 \label{Fourier_psi_P-twisted_in_x_direction_Appendix}
 \end{align}
 with 
\begin{align}
 e^{2\pi i s_x} \chi_{\pm}(\mathsf{s}) = \eta_{\pm}
 e^{2\pi i a_x} 
 \chi_{\pm} (\bar{\mathsf{s}}),
 \quad
 s_x\in 
 \frac{\mathbb{Z}}{2}+a_x,
 \quad
 \eta_{\pm}^2=1,
 \label{relation_chi_pm_s_tilde_s_P-twisted_in_x_direction_Appendix}
 \end{align}
where $\chi_\pm(\mathsf{s})$ are eigen basis of $\mathcal{H}'(\mathsf{s})$, $\vec{u}_{\pm}(\mathsf{s})$ are the corresponding eigenvectors [take the form of (\ref{eigenvectors_u_pm_1_App}) or (\ref{eigenvectors_u_pm_2_App}), up to normalization factors], and the term "2d massless modes" is present if $a_y\in\mathbb{Z}$.
The 2d massless modes are given by the sum of the two terms
\begin{align}
\sum_{s^R_x\in \mathbb{Z}+a_x} 
 e^{  i x \frac{s^R_x}{R_1}} \vec{u}_R(s^R_x)\chi_R(s^R_x),
 \nonumber\\
 \sum_{s^L_x\in \mathbb{Z}+a_x-\frac{1}{2}} 
 e^{  i x \frac{s^L_x}{R_1}} \vec{u}_L(s^L_x)\chi_L(s^L_x),
 \label{zero_modes_psi_P-twisted_in_x_direction_Appendix}
\end{align}
where $\chi_{R,L}(\mathsf{s})$ are eigenbasis of $\mathcal{H}'(s_x, s_y=0)$ and $\vec{u}_{R,L}(\mathsf{s})$ 
are the corresponding eigenvectors in Eq.\ (\ref{eigenvectors_u_RL_App}). 

From the condition (\ref{relation_chi_pm_s_tilde_s_P-twisted_in_x_direction_Appendix}), 
which relates eigen modes with $\mathsf{s}$ and $\bar{\mathsf{s}}$, we only need to take "half" of the degree of freedoms, either modes with $s_y>0$ or with $s_y<0$, when we calculate the trace for the partition functions. The result does not depend on which region for $s_y$ we choose.
From the above discussion, the 2d massive part for fixed $s_y\neq 0$ in the trace 
$\mathrm{Tr}_{\mathscr{P}\mathscr{G}_f^{2a_x},a_y } 
[\mathscr{G}_f^{2(a_{\tau}-1/2)} e^{-2\pi R_0H'}]$ 
is evaluated as 
\begin{align}
 &e^{ 2\pi R_0 \sum_{s_x\in \mathbb{Z}/2+a_x}\varepsilon(\mathsf{s})}
 \nonumber\\
 &\times\prod_{s_x\in \mathbb{Z}/2+a_x} 
 \left|
 1-e^{-2\pi R_0\varepsilon(\mathsf{s})+2\pi i\alpha s_x+2\pi i a_{\tau}}
 \right|^2
 \nonumber\\
 &=
 \Theta_{[2a_x, a_{\tau}] } \left(\tau_{2d}/2; r_{12} s_y\right),
\end{align}
while the 2d massless part (if present) is evaluated as 
\begin{align}
A^R_{[a_x,a_{\tau}]}({\tau_{2d}})A^L_{[a_x-\frac{1}{2},a_{\tau}]}({\tau_{2d}}_{}).
\end{align}
In summary, 
\begin{align}
&Z_{
\mathscr{G}_f^{2a_{\tau}}, \mathscr{P}\mathscr{G}_f^{2a_x},a_y
}
\nonumber \\
&=
\left\{
\begin{array}{l}
\text{const.} \times
A^R_{[a_x,a_{\tau}]}({\tau_{2d}})A^L_{[a_x-\frac{1}{2},a_{\tau}]}({\tau_{2d}}_{})
\\
\displaystyle 
\qquad \times  \prod_{s_y\in\mathbb{Z}^+}  
\Theta_{[2a_x, a_{\tau}] } \left(\tau_{2d}/2; 2r_{12} s_y\right),
\\
\qquad \mbox{for $a_y=0$ (PBC in the $y$-direction)}
\\
\\
\displaystyle 
\text{const.} \times
 \prod_{s_y\in\mathbb{Z}^+-\frac{1}{2}}  \Theta_{[2a_x, a_{\tau}] } \left(\tau_{2d}/2; 2r_{12} s_y\right)
\\
\qquad \mbox{for $a_y=1/2$ (APBC in the $y$-direction)}. 
\end{array}
\right.
\end{align}
%
%

\paragraph{P-twisted partition functions in the ${\tau}$- and $x$-directions}

Finally, we calculate the partition function twisted by P both in the ${\tau}$- and $x$-directions, 
$
Z_{
\mathscr{P}\mathscr{G}_f^{2a_{\tau}},
\mathscr{P}\mathscr{G}_f^{2a_x},a_y 
}
$. 
Using the result from the last section, we now just need to include the additional insertion of the parity operator inside the trace. This can be done by observing that  
\begin{align}
 \mathscr{P}\chi_{\pm}(\mathsf{s})  \mathscr{P}^{-1}
 = \eta_{\pm}\chi_{\pm} (\bar{\mathsf{s}})
 =e^{2\pi i s_x} 
e^{-2\pi i a_{x}} \chi_{\pm}(\mathsf{s})
  \end{align}
for the massive modes ($s_y\neq 0$) and 
\begin{align}
 \mathscr{P} 
 \left[
 \begin{array}{c}
 \chi_{R}(s_x) \\
 \chi_{L}(s_x) 
 \end{array}
 \right] 
 \mathscr{P}^{-1} 
 = \sigma_3
  \left[
 \begin{array}{c}
 \chi_{R}(s_x) \\
 \chi_{L}(s_x) 
 \end{array}
 \right]
 \nonumber
 \end{align}
for the massless modes (where $s_y=0$ as usual).
Then, the 2d massive part for fixed $s_y\neq 0$ in the trace is evaluated as 
\begin{align}
 &e^{ 2\pi R_0 \sum_{s_x\in \mathbb{Z}/2+a_x}\varepsilon(\mathsf{s})}
 \nonumber\\
 &\times\prod_{s_x\in \mathbb{Z}/2+a_x} 
 \left|
 1-e^{-2\pi R_0\varepsilon(\mathsf{s})+2\pi i(\alpha+1) s_x+2\pi i (a_{\tau}-a_x)}
 \right|^2
 \nonumber\\
 &=
 \Theta_{[2a_x, a_{\tau}-a_x ] } 
 \left(\tau_{2d}/2+1/2; 2r_{12} s_y\right),
 \label{massive_modes_partition functions_P-twisted_in_tau_x_directions_Appendix}
\end{align}
while the 2d massless part is evaluated as 
\begin{align}
A^R_{ [a_x,a_{\tau}]}({\tau_{2d}})A^L_{ [a_x-\frac{1}{2},a_{\tau}-\frac{1}{2}]}({\tau_{2d}}).
\end{align}
%
%
In summary, 
\begin{align}
&Z_{
\mathscr{P}\mathscr{G}_f^{2a_{\tau}}, 
\mathscr{P}\mathscr{G}_f^{2a_x},a_y 
}
\nonumber \\
&=
\left\{
\begin{array}{ll}
\text{const.} \times
 A^R_{[a_x,a_{\tau}]}({\tau_{2d}})A^L_{ [a_x-\frac{1}{2},a_{\tau}-\frac{1}{2}]}({\tau_{2d}}_{})
 \nonumber\\
 \displaystyle
\qquad \times  \prod_{s_y\in\mathbb{Z}^+}  
\Theta_{[2a_x, a_{\tau}-a_x] } \left(\tau_{2d}/2+1/2; 2r_{12} s_y\right)
\\
\qquad \mbox{for $a_y=0$ (PBC in the $y$-direction)}, 
\\
\\
\displaystyle 
\text{const.} \times
 \prod_{s_y\in\mathbb{Z}^+-\frac{1}{2}}  \Theta_{[2a_x, a_{\tau}-a_x] } \left(\tau_{2d}/2+1/2; 2r_{12} s_y\right)
 \\
\qquad \mbox{for $a_y=1/2$ (APBC in the $y$-direction)}.
\end{array}
\right.
\label{partition functions_PBC_P-twisted_in_tau_x_directions_Appendix}
\end{align}

\section{Massive modes $\Theta^{i-iv}_{[a_x,a_{\tau}]}(\tau_{2d}; r_{12})$ under $\mathrm{SL}(2, \mathbb{Z})$ transformations}
\label{Appendix_E}

In this Appendix, we discuss how the (products of) massive modes $\Theta^{i-iv}_{[a_x,a_{\tau}]}(\tau_{2d}; r_{12})$, 
defined in Eq.\ (\ref{def_four_Theta}), 
transform under $\mathrm{SL}(2, \mathbb{Z})$ generated by $U'_1$ and $U_2$.
This can be deduced from the modular properties (\ref{massive_theta_func_properties}) of
the massive theta functions with modular parameters 
$\tau_{2d}$, $2\tau_{2d}$, $\tau_{2d}/2$, and $\tau_{2d}/2+1/2$ 
(we denote the mass parameter $m=r_{12}s_y$ in the following equations):
\begin{enumerate}
\item[(i)]
For $\Theta_{[a_x,a_{\tau}]}({\tau_{2d}}; m)$: 
\begin{align}
\Theta_{[a_x,a_{\tau}]}({\tau_{2d}}; m) 
&\overset{U'_1}{\longrightarrow} 
\Theta_{[a_x,a_{\tau}]}\left(-1/\tau_{2d}; m|{\tau_{2d}}|\right) 
\nonumber\\
&= \Theta_{[-a_{\tau}, a_x]}\left({\tau_{2d}}; m\right),
\nonumber \\
\Theta_{[a_x,a_{\tau}]}({\tau_{2d}}; m) 
&\overset{U_2^{-1}}{\longrightarrow} 
\Theta_{[a_x,a_{\tau}]}({\tau_{2d}}+1; m) 
\nonumber\\
&= \Theta_{[a_x,a_x+a_{\tau}]}\left({\tau_{2d}}; m\right);
\end{align}
\item[(ii)]
For $\Theta_{[a_x,a_{\tau}]}(2{\tau_{2d}}; m)$: 
\begin{align}
\Theta_{[a_x,a_{\tau}]}(2{\tau_{2d}}; m) 
&\overset{U'_1}{\longrightarrow} 
\Theta_{[a_x,a_{\tau}]}\left(-2/\tau_{2d}; m|{\tau_{2d}}|\right) 
\nonumber \\
&= \Theta_{[-a_{\tau}, a_x]}\left(\tau_{2d}/2; 2m\right),
\nonumber \\
 \Theta_{[a_x,a_{\tau}]}(2{\tau_{2d}}; m) 
&\overset{U_2^{-1}}{\longrightarrow} 
\Theta_{[a_x,a_{\tau}]}(2{\tau_{2d}}+2; m) 
\nonumber \\
&= \Theta_{[a_x,2a_x+a_{\tau}]}\left(2{\tau_{2d}}; m\right);
\label{mod_trans_massive_theta_func_2}
\end{align}
\item[(iii)]
For $\Theta_{[a_x,a_{\tau}]}({\tau_{2d}}/2; 2m)$: 
\begin{align}
\Theta_{[a_x,a_{\tau}]}\left({\tau_{2d}}/2; 2m\right) 
&\overset{U'_1}{\longrightarrow} 
\Theta_{[a_x,a_{\tau}]}\left(-1/2\tau_{2d}; 2m|{\tau_{2d}}|\right)
\nonumber \\
&    = \Theta_{[-a_{\tau},a_x]}\left(2{\tau_{2d}}; m\right),
\nonumber \\
\Theta_{[a_x,a_{\tau}]}\left(\tau_{2d}/2; 2m\right) 
&\overset{U_2^{-1}}{\longrightarrow} 
\Theta_{[a_x,a_{\tau}]}\left(\tau_{2d}/2+1/2; 2m\right);
\label{mod_trans_massive_theta_func_3}
\end{align}
\item[(iv)]
For $\Theta_{[a_x,a_{\tau}]}({\tau_{2d}}/2+1/2; 2m)$: 
\begin{align}
&\Theta_{[a_x,a_{\tau}]}\left(\tau_{2d}/2+1/2; 2m\right) 
\nonumber\\
&\qquad\qquad\qquad\qquad
\overset{U'_1}{\longrightarrow} 
\Theta_{[a_x,a_{\tau}]}\left(-1/2\tau_{2d}+1/2; 2m|{\tau_{2d}}|\right) 
\nonumber \\
&\qquad\qquad\qquad\qquad= 
\Theta_{[-a_x-2a_{\tau},a_x+a_{\tau}]}\left(\tau_{2d}/2+1/2; 2m\right),
 \nonumber\\
&\Theta_{[a_x,a_{\tau}]}\left(\tau_{2d}/2+1/2; 2m\right) 
\overset{U_2^{-1}}{\longrightarrow} 
\Theta_{[a_x,a_{\tau}]}\left(\tau_{2d}/2+1; 2m\right) 
\nonumber\\
&\qquad\qquad\qquad\qquad= 
\Theta_{[a_x,a_x+a_{\tau}]}\left(\tau_{2d}/2; 2m\right).
\label{mod_trans_massive_theta_func_4}
\end{align}
\end{enumerate}
Therefore,
\begin{align}
&\Theta^{i}_{[a_x,a_{\tau}]}
\overset{U'_1}{\longrightarrow} 
 \Theta^{i}_{[-a_{\tau}, a_x]},
\quad
\Theta^{i}_{[a_x,a_{\tau}]}
\overset{U_2^{-1}}{\longrightarrow} 
 \Theta^{i}_{[a_x,a_x+a_{\tau}]},
\nonumber\\
&\Theta^{ii}_{[a_x,a_{\tau}]}
\overset{U'_1}{\longrightarrow} 
\Theta^{iii}_{[-a_{\tau}, a_x]},
\quad
\Theta^{ii}_{[a_x,a_{\tau}]}
\overset{U_2^{-1}}{\longrightarrow} 
 \Theta^{ii}_{[a_x,2a_x+a_{\tau}]},
 \nonumber\\
&\Theta^{iii}_{[a_x,a_{\tau}]}
\overset{U'_1}{\longrightarrow} 
 \Theta^{ii}_{[-a_{\tau}, a_x]},
\quad
\Theta^{iii}_{[a_x,a_{\tau}]}
\overset{U_2^{-1}}{\longrightarrow} 
 \Theta^{iv}_{[a_x,a_{\tau}]},
 \nonumber\\
&\Theta^{iv}_{[a_x,a_{\tau}]}
\overset{U'_1}{\longrightarrow} 
 \Theta^{iv}_{[-a_x-2a_{\tau},a_x+a_{\tau}]},
\quad
\Theta^{iv}_{[a_x,a_{\tau}]}
\overset{U_2^{-1}}{\longrightarrow} 
 \Theta^{iii}_{[a_x,a_x+a_{\tau}]}.
\end{align}

\section{$\mathrm{SL}(2, \mathbb{Z})$ invariance of the total partition function for $a_y=1/2$}
\label{Appendix_D}

The parity-twisted partition functions for $a_y=1/2$, as computed in Appendix\ 
\ref{Parity twisted partition functions of the surface theory of 
crystalline topological superconductors},
are summarized as follows:
\begin{align}
&
Z_{\mathscr{G}_f^{2a_{\tau}}, \mathscr{G}_f^{2a_x}, a_y=\frac{1}{2}}
 = 
\tilde{\Theta}^{i}_{[a_x,a_{\tau}]}(\tau_{2d}; r_{12}),
\nonumber\\
&
Z_{
\mathscr{P}\mathscr{G}_f^{2a_{\tau}}, \mathscr{G}_f^{2a_x},a_y=\frac{1}{2}
}
= 
\text{const.} \times 
\tilde{\Theta}^{ii}_{[a_x,2a_{\tau}]}(\tau_{2d}; r_{12}),
\nonumber\\
&
Z_{
\mathscr{G}_f^{2a_{\tau}}, \mathscr{P}\mathscr{G}_f^{2a_x},a_y=\frac{1}{2}
}
= 
\text{const.} \times 
\tilde{\Theta}^{iii}_{[2a_x,a_{\tau}]}(\tau_{2d}; r_{12}),
\nonumber\\
&
Z_{\mathscr{P}\mathscr{G}_f^{2a_{\tau}}, \mathscr{P}\mathscr{G}_f^{2a_x},a_y=\frac{1}{2}}
= 
\text{const.} \times 
\tilde{\Theta}^{iv}_{[2a_x,a_{\tau}-a_x]}(\tau_{2d}; r_{12}),
\label{P_twisted_PFs_2}
 \end{align}
where  we have introduced 
$\tilde{\Theta}^{i-iv}_{[a_x,a_{\tau}]}(\tau_{2d}; r_{12})$ as:
\begin{align}
\tilde{\Theta}^{i}_{[a_x,a_{\tau}]}
&=
\prod_{s_y\in\mathbb{Z}^+-1/2}\left[ \Theta_{[a_x,a_{\tau}]}\left(\tau_{2d} ; r_{12}s_y\right) \right]^2,
\nonumber \\
\tilde{\Theta}^{ii}_{[a_x,a_{\tau}]}
&=
\prod_{s_y\in\mathbb{Z}^+-1/2}\Theta_{[a_x,a_{\tau}]}(2\tau_{2d} ; r_{12}s_y),
\nonumber\\
\tilde{\Theta}^{iii}_{[a_x,a_{\tau}]}
&=
\prod_{s_y\in\mathbb{Z}^+-1/2}\Theta_{[a_x,a_{\tau}]}(\tau_{2d}/2 ; 2r_{12}s_y),  
\nonumber \\
\tilde{\Theta}^{iv}_{[a_x,a_{\tau}]}
&=
\prod_{s_y\in\mathbb{Z}^+-1/2}\Theta_{[a_x,a_{\tau}]}(\tau_{2d}/2+1/2 ; 2r_{12}s_y). 
\label{def_four_tildeTheta}
\end{align}
The constant prefactors are again related to the P eigenvalues of the ground states, which can be absorbed to the (redefined) weights 
as we consider the partition sum.

The total partition function is then given by
\begin{align}
&\mathcal{Z}^{tot}_{[a_y=\frac{1}{2}]}(g_{\mathrm{P}})
\nonumber \\
 &=
  \epsilon_{1}\tilde{\Theta}^{i}_{[0,0]}
+\epsilon_{2}\tilde{\Theta}^{i}_{[0,\frac{1}{2}]}
+\epsilon_{3}\tilde{\Theta}^{i}_{[\frac{1}{2},0]}
+\epsilon_{4}\tilde{\Theta}^{i}_{[\frac{1}{2},\frac{1}{2}]}
\nonumber\\
&\quad+
2\Big( 
  \epsilon_{5}\tilde{\Theta}^{ii}_{[0,0]}
+\epsilon_{6}\tilde{\Theta}^{ii}_{[\frac{1}{2},0]}
+\epsilon_{7}\tilde{\Theta}^{iii}_{[0,0]}
+\epsilon_{8}\tilde{\Theta}^{iii}_{[0,\frac{1}{2}]}
\nonumber\\
&\qquad
+\epsilon_{9}\tilde{\Theta}^{iv}_{[0,0]}
+\epsilon_{10}\tilde{\Theta}^{iv}_{[0,\frac{1}{2}]}
\Big).
\end{align}
From the modular properties of $\Theta$ (and thus of $\tilde{\Theta}^{i-iv}$) discussed in Appendix\ \ref{Appendix_E}, 
we can see that $\mathcal{Z}^{tot}_{[a_y=\frac{1}{2}]}(g_{\mathrm{P}})$
can be made $\mathrm{SL}(2,\mathbb{Z})$ (generated by $U'_1$ and $U_2$)
invariant for any  number of Dirac fermion flavors, $N$,
if we choose $\epsilon_i=1$ for all $i$ (more precisely, we just need $\epsilon_2=\epsilon_3=\epsilon_4$ and $\epsilon_5= ... =\epsilon_{10}$).

\bibliography{reference}


\end{document}